\documentclass[aps,prd,twocolumn,showpacs,floatfix,nofootinbib,superscriptaddress]{revtex4-1}
\usepackage{amsmath}
\usepackage{amssymb}
\usepackage{graphicx}
\usepackage{hyperref}
\usepackage{color}
\usepackage{mathtools}
\usepackage{times}
\usepackage{hyperref}
\usepackage{bm}
\usepackage{comment} 

\hypersetup{
    colorlinks=true,
    linkcolor=blue,
    filecolor=magenta,
    citecolor=blue,
    urlcolor=blue,
}

\let\la\undefined
\let\ga\undefined

\newcommand{\DDD}{D}
\newcommand{\bdv}[1]{{\bf #1}}
\newcommand{\mpc}{{\rm Mpc}}
\newcommand{\hmpci}{{h\mpc^{-1}}}
\newcommand{\RA}{\rightarrow}
\newcommand{\obar}{\bar{\rm o}}
\newcommand{\CMB}{\Theta_\ga^\up{obs}}
\newcommand{\ngobs}{n_g^\up{obs}}
\newcommand{\degi}{\de_g}
\newcommand{\deobs}{\de_g^\up{obs}}
\newcommand{\TT}{\mathcal{T}}
\newcommand{\ngtwo}{\Sigma_g^\up{obs}}
\newcommand{\up}[1]{{\rm #1}}

\newcommand{\beeq}{\begin{equation}}
\newcommand{\eneq}{\end{equation}}
\newcommand{\bear}{\begin{eqnarray}}
\newcommand{\enar}{\end{eqnarray}}
\newcommand{\la}{\langle}
\newcommand{\ra}{\rangle}
\newcommand{\nnn}{\nonumber \\}
\newcommand{\nn}{\nonumber}
\newcommand{\AVE}[1]{\left\langle#1\right\rangle}
\newcommand{\pa}{\partial}
\newcommand{\para}{\parallel}
\newcommand{\Dquad}{\qquad\qquad}
\newcommand{\Tquad}{\qquad\qquad\qquad}
\newcommand{\kms}{{\rm km\, s}^{-1}}
\newcommand{\OO}{\mathcal{O}}
\newcommand{\kvec}{\bdv{k}}
\newcommand{\xvec}{\bdv{x}}
\newcommand{\Vang}{\bdv{\hat n}}
\newcommand{\Nang}{\hat{\bm{n}}}
\newcommand{\al}{\alpha}
\newcommand{\ga}{\gamma}
\newcommand{\de}{\delta}
\newcommand{\HH}{\mathcal{H}}   
\newcommand{\rbar}{\bar r}      
\newcommand{\dz}{\delta z}      
\newcommand{\CC}{\mathcal{C}}

\begin{document}

\title{Cosmic Dipoles from Large-Scale Structure Surveys}

\author{Jaiyul Yoo}
\email{jaiyul.yoo@uzh.ch}
\affiliation{Center for Theoretical Astrophysics and Cosmology,
Department of Astrophysics,
University of Z\"urich, Winterthurerstrasse 190,
CH-8057, Z\"urich, Switzerland}
\affiliation{Department of Physics, University of Z\"urich,
Winterthurerstrasse 190, CH-8057, Z\"urich, Switzerland}

\author{Matteo Magi}
\email{matteo.magi@uzh.ch}
\affiliation{Center for Theoretical Astrophysics and Cosmology,
Department of Astrophysics,
University of Z\"urich, Winterthurerstrasse 190,
CH-8057, Z\"urich, Switzerland}

\author{Dragan Huterer}
\email{huterer@umich.edu}
\affiliation{Department of Physics and Leinweber Institute for Theoretical 
Physics, University of Michigan, 450 Church St, Ann Arbor, MI 48109, USA}

\date{\today}

\begin{abstract}
Large-scale structure surveys can be used to measure the dipole
in the cosmic microwave background (CMB), in 
 the luminosity distances inferred from type-Ia supernova observations, 
and in the spatial distribution of galaxies and quasars. 
The measurements of these cosmic dipoles appear to be mutually
inconsistent,
even though they
are expected to indicate the common observer velocity.
This observational tension may represent a significant challenge 
to the standard model of cosmology.
Here we study in detail what contributes to the 
cosmic dipoles from CMB, supernova, and galaxy survey
in the standard $\Lambda$CDM model, though our theoretical model can be
applied beyond the standard model.
 While measurements of the
cosmic dipoles yield the relative velocities between the source samples
and the observer velocity, the motion of the observer is the dominant
contribution in the conformal Newtonian gauge, and the intrinsic velocities
of the samples fall steeply with increasing redshift of the sources.
Hence the cosmic dipoles of CMB, type-Ia supernovae, and galaxies 
should be aligned but can have different amplitudes. 
We also clarify several misconceptions 
that are commonly found in the literature. 
\end{abstract}

\maketitle

\section{Introduction}
\label{intro}

Large-scale structure in the Universe can be probed by tracers such as
galaxies, quasars, type-Ia supernovae (SNIa), 
and cosmic microwave background (CMB)
 temperature 
anisotropies, and these large-scale structure tracers form an effective fluid
that is described in terms of its density, velocity, and pressure. 
The evolution of such cosmic fluids 
is governed by gravity, and the current standard model in cosmology
is the $\Lambda$CDM model with Einstein's theory of general relativity.
Hence, a test of the standard model in cosmology by using the mutual
relation and time evolution of the cosmic fluids 
is one of the ultimate goals in large-scale structure surveys.
While the first moment (or the density fluctuations) and its
two-point correlation function (or the power spectrum) have received
most attention in large-scale structure surveys
(see, e.g., \cite{TEEIET06,AUBAET15,DESIbao24}), the second moment (or peculiar
velocities) and its spatial correlations also contain crucial cosmological
information such as the logarithmic growth rate derivable from the
redshift-space distortion \cite{KAISE87}.

In particular, the peculiar velocities of the tracers of
large-scale structure measured
at the position of the observer leave unique
dipolar patterns, and their dipole measurements
can be used to test the standard model of cosmology \cite{ELBA84}.
The observed CMB sky is nearly isotropic,
 with anisotropies only at the level of $10^{-5}$, but this
isotropic sky temperature 
map is obtained only after subtracting the dipole pattern
with an amplitude
100 times larger than the rms fluctuation amplitude. This dipole moment was
measured in various CMB satellite missions 
\cite{KOLIET93,HIWEET09,PLANCKdop13,PLANCKover18}, culminating
with  the sub-percent level measurement precision 
in the Planck survey \cite{PLANCKover18}.
Large-scale galaxy surveys also produce sky maps of galaxies, quasars,
or any other cosmological tracers  (collectively referred to as galaxies) 
primarily as a function of angular position,
with additional information of flux, redshift, and so on
(see, e.g., \cite{YOADET00,CODAET01}). In the same way as in the CMB sky map,
the peculiar velocities of these galaxy samples
imprint the dipolar patterns in the galaxy sky maps,
and the dipole moments can be measured for individual galaxy samples
in large-scale galaxy surveys.
In practice, however, it is more difficult to measure the dipole moments
in galaxy surveys due to the limited sky coverage than in CMB surveys with
an almost full-sky coverage. 
The dipole moment of a galaxy survey was first measured 
\cite{BLWA02,SINGA11,GIHU12,RUSC13,TIKOET15,TINU16} in the 
 NRAO VLA Sky Survey (NVSS; \cite{COCOET98}).
Other large-scale galaxy surveys in the following years were used
to measure the dipole moments of various cosmological sources
\cite{GIHU12,BEMASA18,SINGA19,SEHAET21,TISCET24,SINGA24} with increasing
measurement precision.

The dipole patterns in the large-scale structure surveys mainly arise
from the peculiar motion of the observer, and hence the direction
and the amplitude of the peculiar motion in responsible for the dipole 
measurements should be identical to those found in the CMB sky map.
This cosmological test using the dipole measurements from various large-scale
structure surveys is known as the Ellis-Baldwin test \cite{ELBA84}.
From the inception of the Ellis-Baldwin test,
large-scale galaxy surveys tend to report
that the dipole direction is largely aligned with the CMB dipole direction,
but the inferred amplitude is larger by a factor of few than in the CMB
sky map (see \cite{SEHAET25} for a recent review). Such discrepancy has worsened
in the recent years, and in particular the precision measurement of the
dipole moment for the quasar sample from the WISE survey \cite{WREIET10}
puts the discrepancy at the 5-$\sigma$ level  \cite{SEHAET21},
calling for close investigations of the standard model of cosmology
\cite{PEEBL25}.

Here we discuss in detail what contributes to the
dipole moments in large-scale structure surveys, and we clarify
several misconception in the literature. For numerical computation,
we adopt the standard $\Lambda$CDM model with cosmological parameters
obtained in the Planck cosmological parameter analysis \cite{PLANCKover18}
and use
the Boltzmann solvers
(\textsc{CAMB} \cite{LECHLA00} and \textsc{CLASS} \cite{CLASS}) 
to compute our theoretical predictions. Though our analytical and numerical
calculations are based on the standard cosmological model, they can be
readily applied to the beyond-the-standard models.
In Section~\ref{subtle} we discuss subtleties associated with relative motion
and the Ellis-Baldwin test. Then in Sections~\ref{Dcmb}$-$\ref{LSS},
we discuss the dipole moments from the CMB temperature anisotropies
 (\S~\ref{Dcmb}), the supernova surveys (\S~\ref{DSN}), and the galaxy
surveys (\S~\ref{LSS}). We discuss our findings in Section~\ref{summary}.
The details of our calculations can be found in Appendices.

\section{Subtleties in cosmic dipoles}
\label{subtle}
In a homogeneous and isotropic universe (or a background universe),
all observers on a time-like geodesics or observers following 
any fluids such as the matter and the photon fluids
are at rest, and this rest frame is shared by all the observers
(hence the {\it cosmic rest frame}).
A coordinate system that describes the homogeneous and isotropic
universe is uniquely determined up to the rotation and translation of the
space. In a real universe with inhomogeneities, however,
there is no unique way to fit a background homogeneous and isotropic
universe (or a coordinate system) to the real universe. This fitting process
amounts to a gauge choice \cite{ELST87,ELBR89},
and all gauge choices are on the same
footing in general relativity with diffeomorphism symmetry.
One can therefore choose {\it any} coordinate system 
(which represents a background
homogeneous and isotropic universe) and work with perturbations (or the
difference between the real universe and the chosen background universe 
at a given spacetime position). 
These perturbations are hence gauge-dependent (or coordinate dependent), 
and their interpretations also depend on the choice of gauge (see, e.g.,
\cite{BARDE80,KOSA84,MUFEBR92}).

Cosmic dipoles have been measured in the past by using
the CMB temperature anisotropies
\cite{KOLIET93,HIWEET09,PLANCKdop13,PLANCKover18}
and other large-scale structure probes such as supernovae \cite{BODUKU06},
quasars \cite{SEHAET21},  and galaxies 
\cite{GIHU12,YOHUGIET14,BEMASA18,SINGA19,SISCSC21,TISCET24,SINGA24}.
The fact that we measure dipoles
is a manifestation that we live in an inhomogeneous universe, not in
a smooth
background universe, and the subtleties associated with gauge choice
in interpreting the cosmic dipole measurements are inevitable.
Here we clarify  common misunderstanding in the literature
associated with the cosmic
dipole measurements. However, keep in mind that these subtleties 
only exist
in the theoretical interpretations, not in the measurements themselves.

\subsection{Relative velocities}
In special relativity, the absolute velocity of any object
has no physical meaning, and only the velocity 
relative to an observer (or relative velocity) is physically meaningful.
In general relativity, a comparison of two velocities is meaningful only
at the same position, as a parallel transport of velocities to a different
position is path-dependent. So, the relative velocity from two different
spacetime positions has no physical meaning either.
In the context of cosmological perturbation theory, individual
velocities are gauge-dependent, and they change their values depending on
a coordinate choice. Hence, these individual velocities cannot be
associated with any of the dipole measurements, and 
the only physically meaningful velocity is 
the gauge-invariant relative velocity at the same spacetime position.
The observed dipoles from large-scale
structure surveys are indeed measurements of the  relative velocities
at the same spacetime position, i.e.,
\beeq
\label{EQ:RELVEL}
\bm{v}_\up{CMB}=\bm{v}_\ga-\bm{v}_\up{o}~,\qquad\qquad 
\bm{v}_\up{LSS}=\bm{v}_g-\bm{v}_\up{o}~,
\eneq
where $\bm{v}_\ga$, $\bm{v}_g$, $\bm{v}_\up{o}$ 
are the velocities of the CMB photon fluid, the galaxy sky map, 
and the observer, and $\bm{v}_\up{CMB}$,
$\bm{v}_\up{LSS}$ are the velocities 
inferred by using the dipole measurements from the CMB anisotropies and
the large-scale structure probes (quasars and supernovae are collectively
referred to as galaxies with~$\bm{v}_g$). We clarify the meaning 
of~$\bm{v}_g$ at the observer position in Section~\ref{subtleC}.

Note that while the individual velocities $\bm{v}_\ga$, $\bm{v}_
g$, $\bm{v}_\up{o}$ are gauge-dependent,
the relative velocities~$\bm{v}_\up{CMB}$ 
and~$\bm{v}_\up{LSS}$ are gauge-invariant, as the gauge
transformation of the two individual velocities
at the same position is cancelled.
Hence, the relative velocities are physically meaningful, but
the individual velocities are of less physical significance.
For instance, popular Boltzmann codes 
(\textsc{CMBFAST} \cite{SEZA96}, \textsc{CAMB} \cite{LECHLA00}, 
\textsc{CLASS} \cite{CLASS}) adopt the synchronous and dark-matter comoving
gauge, i.e., $\bm{v}_\up{o}^\up{sync}\equiv0$,
if we assume that the observer is (co-)moving
together with dark matter. With this gauge condition, 
the observer is  not moving at all, and the relative velocities
from the dipole measurements are $\bm{v}_\up{CMB}=\bm{v}_\ga^\up{sync}$ and 
$\bm{v}_\up{LSS}=\bm{v}_g^\up{sync}$, where we used the superscript to indicate
that the quantities are evaluated in the synchronous-comoving gauge.
Of course, the values of~$\bm{v}_\up{CMB}$ and~$\bm{v}_\up{LSS}$ are
independent of gauge choice.

Viewed this way, the observed dipoles from the CMB and the galaxy
surveys are  not directly related to each other, as they are two different
relative velocities
with only the observer velocity in common. It is, however, possible that they
{\it happen} to be similar to each other --- For example, if the observer
velocity is the dominant contribution 
in a certain gauge, say, the conformal Newtonian (cN) gauge:
\beeq
\label{inequal}
|\bm{v}_\up{o}^\up{cN}|\gg|\bm{v}_\ga^\up{cN}|~,\qquad\qquad
|\bm{v}_\up{o}^\up{cN}|\gg|\bm{v}_g^\up{cN}|~,
\eneq
we could then obtain
\beeq
\label{obs}
\bm{v}_\up{CMB}\simeq\bm{v}_\up{LSS}\simeq-\bm{v}^\up{cN}_\up{o}~.
\eneq
Note that a gauge choice is a matter of convenience and any gauge choices
are equivalent, such that with other gauge conditions the numerical values
of individual velocities~$\bm{v}_\ga$, $\bm{v}_g$, $\bm{v}_\up{o}$ will
be different, but the numerical values of the relative velocities 
($\bm{v}_\up{CMB}$, $\bm{v}_\up{LSS}$) are invariant, i.e.,
\beeq
-\bm{v}_\up{o}^\up{cN}\simeq\bm{v}_\ga^\up{sync}\simeq\bm{v}_g^\up{sync}~,
\eneq
if Eq.~\eqref{inequal} is valid. In the synchronous gauge,
the observer and the dark matter velocities are set zero by a gauge choice,
which then shifts all the other velocities, 
compared to those in the conformal Newtonian gauge.
It is now clear that the peculiar motion of the observer quoted in 
the literature 
indeed corresponds to $\bm{v}_\up{CMB}$ and $\bm{v}_\up{LSS}$, 
not $\bm{v}_\up{o}$ itself. But with inequalities in Eq.~\eqref{inequal}, 
the observed peculiar motion~$\bm{v}_\up{o}^\up{cN}$ in the conformal
Newtonian gauge inferred from the dipole measurements would 
then be the same as in Eq.~\eqref{obs}.

\subsection{No common cosmic rest frames}
Furthermore, Eq.~\eqref{EQ:RELVEL} makes it clear that
the rest frames of CMB and galaxies, in which an observer would see no dipole
(or no spatial energy flux), are  not unique, but position-dependent,
as all individual velocities are a function of spacetime position~$x^\mu$.
The rest frames can be unique and shared by all the observers and 
sources in a background
universe with perfect homogeneity and isotropy, but in a real universe with 
inhomogeneities no common rest frames exist. In this respect,
there exist  no fundamental observers or ``comoving observers'' either, 
who by implicit definition see no dipole of the CMB, galaxies, and any other
sources. 

 In particular, the term ``comoving'' is often misused in the literature 
this way, but a comoving observer simply implies that the observer is 
``co-moving'' with a certain fluid, such that the observer sees  no
dipole of that fluid but the observer can measure dipoles of other fluids. 
A comoving observer is also used to refer to an observer stationary with
respect to comoving coordinates. A better name for this observer would be
a coordinate observer, because, in contrast with
 an observer co-moving with a fluid,
a coordinate observer simply represents a coordinate system, not associated
with any physical quantity. For instance, a normal observer~$n^\mu$ in the
Arnowitt-Deser-Misner formalism (ADM; \cite{ADM}) is an example of a
coordinate-dependent observers (see \cite{YOO14b} for discussion of various
observers and gauge choices).

Because there is only one observer (us) at our position,
there is an approximate rest frame
in which we see no dipoles from CMB, galaxies, and any other sources,
as long as the inequality in Eq.~\eqref{inequal} is valid for all the 
large-scale structure probes. In this regard, this rest frame 
is unique for all sources, relative to the observer at our position.
We show that in the standard cosmology
the inequality is satisfied
for sources at redshift $z\gg0.5$, but not in the local neighborhood.
Furthermore, note that these individual velocities $\bm{v}_\ga^\up{cN}$,
$\bm{v}_g^\up{cN}$ in the conformal Newtonian
gauge are never zero in an inhomogeneous universe even at high redshift,
but just numerically small compared to~$\bm{v}_\up{o}^\up{cN}$.
Hence the ``cosmic rest frame'', even when confined to the only one observer
at one spacetime position, is only an approximate one.

\subsection{Intrinsic velocities and dipoles}
\label{subtleC}
What are the individual velocities $\bm{v}_\ga$, $\bm{v}_g$ of the CMB and
galaxies at the observer position? To a good approximation, CMB photons we
measure today originate from the last scattering surface at~$z_\star$, such
that the CMB sky map has much information about the Universe when 
it was just 370 thousand years old \cite{PLANCKover18}.
However, the CMB sky map itself is indeed a photon fluid 
{\it today at the observer position} (in fact
more than a fluid, as it possesses higher multipoles), 
and that is why we can measure it! Collectively as a fluid, the CMB photons 
form a CMB fluid velocity~$\bm{v}_\ga$ at the observer position
in a given choice of coordinate, while the individual photons move
at the speed of light.

In the same way, galaxy surveys provide the galaxy number density 
$\ngobs(z,\Nang)$
as a function of the observed redshift~$z$ and angle~$\Nang$, or one
can construct the galaxy sky map~$\ngtwo(\Nang)$ by projecting along the
line-of-sight direction as in the CMB temperature map.
It is well known that the density fluctuation \cite{KAISE84}
and the redshift-space distortion \cite{KAISE87}
 are the dominant contributions at redshift~$z$
of the source galaxies, but again the observed galaxy maps (either 
three-dimensional or projected two-dimensional) are indeed
physical observables {\it today} at  {\it the observer position}, which defines
a velocity~$\bm{v}_g$ of the observed source galaxies
at the observer position in a given choice of
coordinate.

These velocities $\bm{v}_\ga^\up{cN}$ and $\bm{v}_g^\up{cN}$ 
 are often in the literature
referred to as the {\it intrinsic dipoles}, as they contribute to the dipole
measurements. However, as we elaborate below, the ``intrinsic dipole''
is misleading in many ways. For example, 
if we boost the observed CMB anisotropies with the relative
velocity $\bm{v}_\up{CMB}$ we obtain from the CMB dipole measurements, 
there will be  no further dipole anisotropies in the boosted frame
(i.e., the rest frame of CMB indeed), rather than we see the residual
dipole from the {\it intrinsic dipole} (see, e.g., \cite{YAPI17}
for the discussion
of the intrinsic dipole of CMB temperature anisotropies). These
velocities $\bm{v}_\ga^\up{cN}$ and $\bm{v}_g^\up{cN}$ are in a sense
{\it intrinsic} to the observed sources, such that they may better be
referred to as {\it intrinsic velocities} of the sources 
(at the observer position). We clarify further in Sections~\ref{Dcmb}
and~\ref{LSS} with equations for~$\bm{v}_\ga^\up{cN}$ and~$\bm{v}_g^\up{cN}$.

\subsection{Bulk velocity flows}
Measurements of peculiar velocities are in general difficult, but there exist
a few ways to measure the peculiar velocities in the local neighborhood
(see, e.g., \cite{WAFEHU09,KAATET08} and the references therein). 
In particular, measurements of the luminosity distance
from supernova observations in the local neighborhood provide a way to
directly estimate the peculiar velocity of their host galaxies.
With these measurements of the peculiar velocities, the bulk velocity
flow can be measured \cite{WAFEHU09,FEWAHU10,TUHUET12,WAALET23}
by averaging the peculiar velocities over the volume with weights chosen
to optimize the signal-to-noise ratio. Anomalously large amplitudes
of the bulk flows \cite{WAFEHU09,KAATET10,WAALET23}
pose another challenge in the standard $\Lambda$CDM cosmology,
and there have been attempts \cite{MAGOFE11}
 to explain the observations
in terms of large-scale or super-horizon scale fluctuations.

As we show in Section~\ref{DSN}, these measurements of the peculiar
velocities are measurements of the relative velocity $\bm{v}_\up{bulk}:=
\bm{v}_\up{s}-\bm{v}_\up{o}$ of the source and the
observer, not the individual peculiar velocity~$\bm{v}_\up{s}$ of the sources. 
Here again note that the individual velocity~$\bm{v}_\up{s}$ and the observer
velocity~$\bm{v}_\up{o}$ alone are gauge-dependent and they are not physical.
Further averaging over the volume of~$\bm{v}_\up{bulk}$ results in
the relative velocity between the observer and the averaged sources.
It is then evident that the presence of any fluctuations on scales larger
than the averaged volume or the separation between the sources and the observer
cannot affect the relative velocity measurements (hence the bulk flow),
as both the velocities are affected by such large-scale fluctuations
(see \cite{BIYO17,MIYOMA23,MAYO23} for the absence of infrared effects on the
luminosity distance). The only way to avoid this conclusion is the
existence of fluctuations or forces that treat the sources and the observer
differently, as opposed to gravity that affects equivalently
 all the matter components at the same position.

\subsection{Cosmological principle on large scales}
The cosmological principle states that the Universe is homogeneous and
isotropic on large scales (see \cite{WEINB72,PEEBL93}). This statement can be
confusing in many ways: at what scale does the Universe become homogeneous and 
isotropic? how do we average the Universe to recover the homogeneity and
isotropy? to what level of fractional deviation does the Universe become
homogeneous and isotropic? The cosmological principle is often invoked 
in the literature for the Ellis \& Baldwin test. 
Here we clarify a few subtleties associated with the cosmological principle
in the literature.

 First, the cosmological principle provides a way to derive
the solution in the nonlinear Einstein equation, i.e., the Robertson-Walker
metric and the Friedmann equation, both of which form a background solution.
In a background universe, the universe is homogeneous and isotropic 
on {\it all scales}. With a background solution, we model the real Universe with
inhomogeneities \cite{BARDE80,ELST87,ELBR89},
by treating the deviations from a background solution 
as perturbations to the background universe we chose
(or cosmological model parameters), which allows us to derive perturbative
solutions to the Einstein equation in the real Universe. Hence the
cosmological principle is to some degree equivalent to a background solution
and tests of the cosmological principle in practice
should be performed as tests of our cosmological model predictions that
include not only the background evolution but also the perturbation
calculations  such as  CMB power spectrum and so on. Note that with 
inhomogeneities the real Universe is not homogeneous or isotropic on any
scales up to the level of perturbations.

Averaging in cosmology in a hyper-surface of a constant time
is known to be gauge-dependent \cite{LAREN09,BRLACO13},
i.e., one obtains
different results, depending on our choice of coordinates, along which we
average. Furthermore, beyond the linear order in perturbations we never
restore the background solution upon averaging, though the so-called
back-reaction is shown to be small in the conformal Newtonian gauge
(see, e.g., \cite{FLWA96,GRWA14,ADCLET15}).
As coordinate averages in a hyper-surface are unphysical and gauge-dependent, 
we need to provide theoretical descriptions that naturally match the 
observational procedure, i.e., averages over the light cone in terms of
the observed angle and redshift 
(see, e.g., \cite{GAMAET11,YOMIET19,MIYOET20}).

Of particular interest in this work is
 the cosmic dipole measurements, i.e., averages over the observed angular
direction of CMB temperature anisotropies and galaxy samples.
As we show in Sections~\ref{Dcmb}$-$\ref{LSS}, these large-scale
structure probes have effective distance~$\rbar_z$ to the sample at
redshift~$z$, and
the dipole measurements yield the fluctuations such as the matter density
fluctuation for galaxy clustering at $k\sim1/\rbar_z$, such
that by using galaxy samples at higher redshift or CMB temperature anisotropies
we probe larger scales, at which the fluctuations are smaller than our
peculiar motion at $\rbar=0$, while at lower redshift ($z\ll 0.5$)
the dipole measurements can be dominated by fluctuations other 
 than our peculiar motion. The statement
that once averaged over large scales we can recover the homogeneity
and isotropy and hence the dipole measurements yield our motion
is an approximate, but not an accurate description.

\section{Dipole from CMB anisotropies}
\label{Dcmb}
The sky of the cosmic microwave background is nearly isotropic at
the level of $10^{-5}$, but our motion relative to the CMB rest frame 
(or the relative velocity $v_\up{CMB}\sim10^{-3}$) induces a dipole
anisotropy an order-of-magnitude larger in amplitude than anisotropies 
on other  angular multipoles. The CMB dipole was first discovered
\cite{CONKL69} in late 1960s and later measured with higher precision by
satellite experiments \cite{KOLIET93,HIWEET09,PLANCKdop13,PLANCKover18}.
The relative motion between the observer and the CMB rest frame generates 
not only the Doppler boost of the CMB temperature 
\cite{PEEBL71,KAKN03,PLANCKdop13},
but also the aberration of the photon directions. The aberration effect
can be measured by the change in the angular power spectrum on large
multipoles \cite{PLANCKdop13,CHVA02,JECHET14},
but its impact is small, compared to the change in
the dipole. Here we focus only on the computation of the
CMB dipole ($l=1$).

CMB temperature anisotropies on low angular multipoles can be accurately
described by a simple analytic formula \cite{SAWO67},
obtained by assuming 1)~the
tight coupling of the baryon and the photon fluids before the recombination,
2)~an instantaneous recombination at~$T_\star$ set by atomic physics,
3)~the free-streaming of photons after the recombination. The observed
CMB temperature anisotropies along the observed angular direction~$\Nang$
are at the linear order in perturbations  \cite{BAYO21}
\bear
\label{CMB}
\CMB(\Nang)&=&\Theta_\star-\dz_\star
=\Theta^\up{cN}_\star+\HH_0 v_{\obar} +\psi_\star-\psi_{\obar}\nnn
&&
+(\pa_\para v)_\star-(\pa_\para v)_{\obar}
+\int_0^{\rbar_\star}d \rbar~(\psi-\phi)'~,~~~~
\enar
where $\Theta$ is the (dimensionless) photon temperature fluctuation,
$\dz$ is the fluctuation in the observed redshift defined by
$T_\star/T(\Nang)=:1+z_\up{obs}(\Nang)=(1+z_\star)(1+\dz)$, and the
subscripts indicate that the field is evaluated at~$\eta_\star$ 
along the light cone or at the observer position~$\obar$.
In the second equality, we expressed the equation in the conformal
Newtonian gauge.
The observed CMB temperature anisotropies are
essentially the temperature fluctuation~$\Theta_\star$ at the recombination and
the relativistic effects~$\dz_\star$ along the propagation such as the
Sachs-Wolfe effect~$\psi$, the line-of-sight Doppler effect $\pa_\para v$,
and the integrated Sachs-Wolfe effect  \cite{SAWO67} 
along the line-of-sight direction to the decoupling $\rbar_\star=13.875$~Gpc
at $z_\star=1088$ (see Appendix~\ref{details} for our notation convention).
  Only the scalar contributions are considered here.
  The analytic formula based on the
aforementioned approximations has been widely used in the
 literature to understand
the CMB temperature anisotropies, and the analytic formula matches perfectly 
the numerical output from the Boltzmann codes on low angular multiples
\cite{SAWO67,SELJA94,HUSU95,ZISC08,ELDU18,BAYO21}. 
The gauge-invariance of the monopole fluctuation in the observed CMB
temperature anisotropies was first proved in \cite{BAYO21} by using
 Eq.~\eqref{CMB}.

Given the analytic expression in Eq.~\eqref{CMB} and its
angular decomposition with the multipole moments~$a_{lm}$
\beeq
\label{MULTIPOLE}
\CMB(\Nang)=\sum a_{lm}Y_{lm}(\Nang)~,
\eneq
the dipole moment~$a_{1m}$ and its power spectrum~$C_1^\up{CMB}$ 
can be readily computed as
\beeq
\label{CMBdipolepow}
C_1^\up{CMB}=\left\la|a_{1m}|^2\right\ra=4\pi\int d\ln k~\Delta_{\cal R}^2(k)~
\left|{\cal T}_1^\up{CMB}(k)\right|^2~,
\eneq
with the dipole transfer function \cite{BAYO21}
\bear
\label{CMBdipole}
&&\hspace{-20pt}
{\cal T}_1^\up{CMB}(k)=
\left[\TT_{\Theta^\up{cN}}(\eta_\star)+\TT_{\psi}(\eta_\star)\right]
 j_1(k\rbar_\star)+k\TT_{v_\ga^\up{cN}}(\eta_\star)j_1'(k\rbar_\star) \nnn
&&
-\frac13k\TT_{v_m^\up{cN}}(\eta_{\obar})+\int_0^{\rbar_\star}d\rbar
~\left[\TT_{\psi'}(\eta)
-\TT_{\phi'}(\eta)\right]j_1(k\rbar)~,~~~~~
\enar
where~$\TT_{\de p}$ is the transfer function (we suppressed the $k$-dependence)
for the individual perturbation variable~$\de p$ defined as 
$\de p(\bm{k},\eta)=\TT_{\de p}(k,\eta){\cal R}(\bm{k})$
in terms of the comoving-gauge curvature perturbation~${\cal R}(\bm{k})$ in the
initial condition, and $\Theta^\up{cN}$, $v_\ga^\up{cN}$, $v_m^\up{cN}$ represent
the CMB temperature fluctuation, the photon velocity potential, and the
matter velocity potential in the conformal Newtonian gauge.
The derivation of the dipole transfer function is 
presented in Appendix~\ref{details}.
Here we assumed that the observer motion is the same as the matter motion.

Equation~\eqref{CMBdipole} reveals that the contribution from the observer
motion~$v_m^\up{cN}$ has  no spherical Bessel function, 
as it arises from the observer position ($\rbar\equiv0$). 
The other contributions arising
from the position at~$\rbar_\star$ contain not only their intrinsic angular
momentum (e.g., scalar for~$\psi$, vector for~$v_\ga^\up{cN}$), but also
the orbital angular momentum from $\exp[i\kvec\cdot\xvec]$. The presence
or the absence of the spherical Bessel function is critical in determining
which component contributes the most to the dipole transfer function,
as the presence of the spherical Bessel function suppresses the
individual transfer functions: $j_l(x)\approx1/x$ at large~$x$.
This structure in the dipole transfer function is identical to the
dipole transfer function for the luminosity distance fluctuations
in Section~\ref{DSN} and for the galaxy surveys in Section~\ref{LSS}.

The dipole transfer function~${\cal T}_1^\up{CMB}$
in the conformal Newtonian gauge has 
the contribution ($-k\TT_{v_m^\up{cN}}/3$)
from the peculiar motion of the observer (or matter), and
the remaining contributions from everything else in Eq.~\eqref{CMBdipole}
(or everything in Eq.~[\ref{CMB}] except $\pa_\para v$ at
the observer position) can be lumped together to define
the {\it intrinsic velocity}~$\bm{v}_\ga^\up{cN}$
of the CMB photon fluid at the observer position today:
\beeq
\label{intCMB}
{\cal T}^\up{CMB}_1(k)=:\frac13k\bigg[\TT_{v_\ga^\up{cN}}(k,\eta_{\obar})
-\TT_{v_m^\up{cN}}(k,\eta_{\obar})\bigg]~,
\eneq
where the numerical factor~1/3 in~${\cal T}_1^\up{CMB}$ 
arises due to equal contributions
from each $m$-modes of the dipole $(l=1)$
and also note that the velocity potential is related
to the velocity vector as $\bm{v}=-\nabla v$.
The intrinsic velocity potential~$v_\ga$ of the CMB photon fluid
at the observer position today defined above indeed gauge
transforms like a velocity
potential, and hence the difference in two velocity potentials at the
same position (or the relative velocity) is gauge-invariant.
As discussed in Section~\ref{subtleC}, this shows that the dipole power
is just a measurement of the relative velocity of the observer and
the CMB fluid today at the observer position.

This argument is further borne out by the fact that the dipole is related
to the spatial energy flux of the CMB photon fluid measured by the observer.
Measurements of CMB temperature anisotropies $\Theta_\ga^\up{obs}(\Nang)$
yield the full information about the distribution function  at the observer 
position 
\beeq
f^\up{obs}_\ga(\Nang)=-{d\bar f\over d\ln q}~\Theta_\ga^\up{obs}(\Nang)~,
\eneq
and thereby the observed energy-momentum tensor of the CMB photon fluid
\beeq
T^{ab}_{\up{obs},\ga}
=2\int d^3E_\ga~{E^a_\ga E^b_\ga\over E_\ga}f^\up{obs}_\ga(\Nang)~,
\eneq
where $\bar f$ is the Planck distribution, $q=aE_\ga$ is the 
photon comoving momentum, $E_\ga^a=E_\ga(1,-\Nang)$, and $a,b=t,x,y,z$
are the internal coordinates in the observer rest frame with
the Minkowski metric (see \cite{MIYO20} for the tetrad formalism).
In particular, the spatial energy flux measured
in the rest frame of the observer
\beeq
T^{0i}_{\up{obs},\ga}=s^i=(\bar\rho+\bar p)_\ga v^i_\up{CMB}~,
\eneq
is related to the dipole moment~$a_{1m}$ as
\beeq
v^i_\up{CMB}=\sqrt{3\over4\pi}\left({a_{1,-1}-a_{11}\over\sqrt2},~
i{a_{1,-1}+a_{11}\over\sqrt2},~a_{10}\right)~,
\eneq
where the multipole coefficients are defined in Eq.~\eqref{MULTIPOLE}.

In the $\Lambda$CDM model the contribution of the intrinsic 
velocity~$\bm{v}_\ga^\up{cN}$ to the dipole power 
is negligible ($\sim10^{-4}$), compared to the
kinematic dipole contribution from~$\bm{v}_\up{o}^\up{cN}$. 
However, $\bm{v}_\ga^\up{cN}$
is non-vanishing and differs at each position, such that  no unique
rest frame of CMB shared by all the observers exists.
Also note that what we can obtain from the CMB dipole measurements is only the
relative velocity of the observer and the CMB photon fluid, not the
observer motion~$\bm{v}_\up{o}^\up{cN}$ in the conformal Newtonian gauge.
Hence, when we boost the observed CMB anisotropies with the relative
velocity we obtain from the CMB dipole measurements, 
there exist no further dipole anisotropies from the intrinsic velocity,
i.e., the rest frame indeed, and the {\it intrinsic} dipole is a misnomer.

\begin{figure}[t]
\centering
\includegraphics[width=0.5\textwidth]{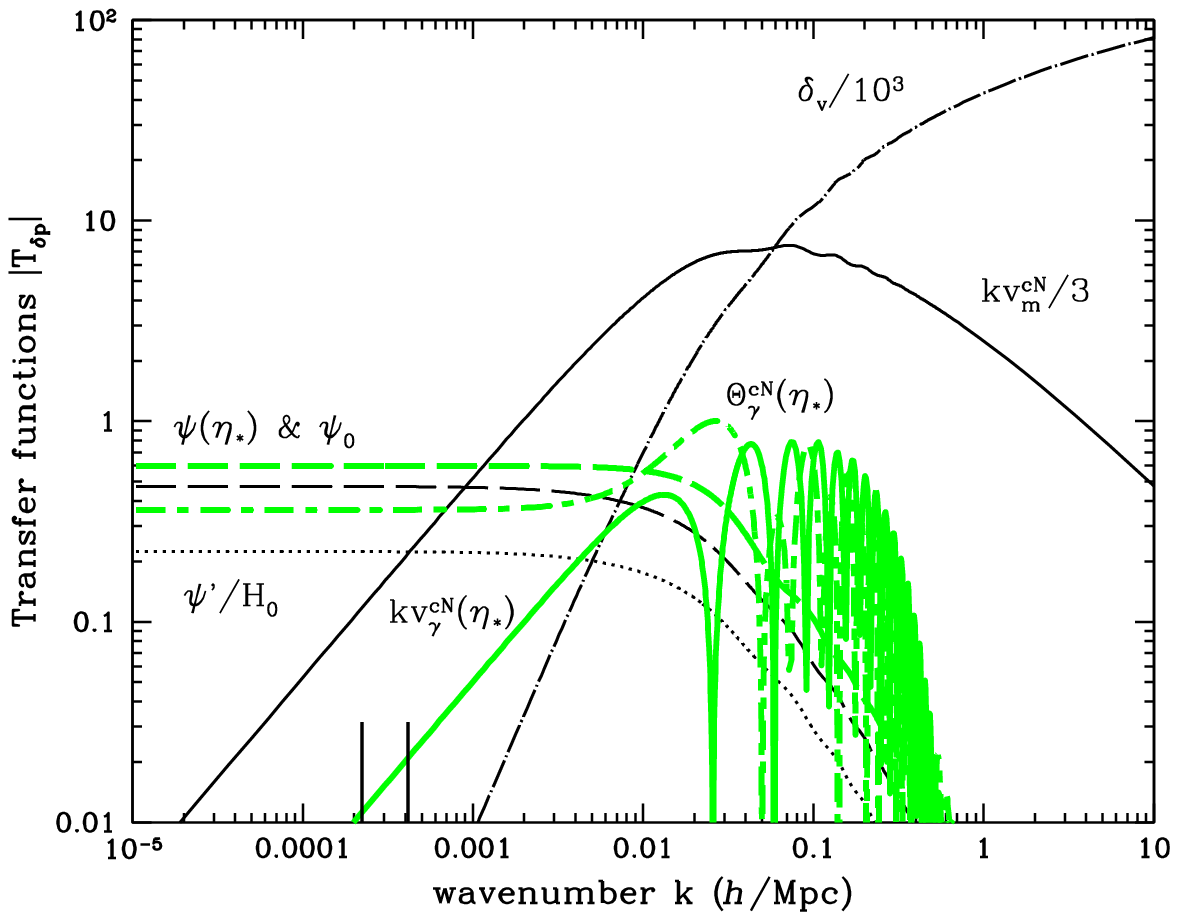}
\caption{Transfer functions~$\TT_{\de p}$
for the individual contributions to the dipole
power spectra in Eqs.~\eqref{CMBdipole}, \eqref{TFSN}, \eqref{TFLSS}.
Green curves show the transfer functions at the decoupling epoch~$z_\star$, 
and they oscillate
on small scales, while the other curves show the transfer functions at $z=0$. 
The gravitational potential changes little over time ($\psi_\star$, $\psi_o$),
but the growth of the velocity potential is significant. None 
of the contributions to the transfer function is comparable
to the matter density fluctuation~$\delta_v$ in the comoving gauge (dot-dashed).
Two short vertical lines indicate the peak
positions $x=2.08$ and $x=3.87$ for~$j_1(x)$ and~$j_1'(x)$ with 
$x=k\rbar_\star$, beyond which the transfer functions are suppressed by the
spherical Bessel function, except the contribution of the observer motion 
(solid) without the spherical Bessel function. Note that only the individual
transfer functions are plotted here, not including the spherical Bessel
function contribution. At lower redshift,
the peak positions are shifted to higher wave numbers.}
\label{Fig:TF}
\end{figure}

Figure~\ref{Fig:TF} shows the transfer functions~$\TT_{\de p}$ in 
Eq.~\eqref{CMBdipole}. The gravitational potential contributions~$\psi$ 
today (dashed) and at the recombination (green dashed)
are constant on large scales, but decay on small scales. Its time evolution
from the recombination epoch until today is very small, mostly arising
from  the epoch of $\Lambda$-domination, as shown by $\psi'/H_0$ (dotted). 
The temperature
fluctuation~$\Theta^\up{cN}$ (green dot dashed)
in the conformal Newtonian gauge is 2/3
of the gravitational potential on large scales, but exhibits acoustic
oscillations on small scales. The velocity contribution~$kv_\ga^\up{cN}$
of the photon fluid at the recombination (green solid) falls as~$k$ on large
scales with the velocity potential being constant, while it also oscillates
on small scales. The velocity contribution~$kv_m^\up{cN}/3$ of the observer
today (solid) has no significant oscillations as assumed to be the
matter component, and it is significantly larger in amplitude as it has
grown from the recombination epoch until today by a factor 28,
or the ratio of $\HH fD$ at two epochs, where~$\HH$ is the conformal
Hubble parameter, $D$~is the growth factor, and~$f$ is the logarithmic
growth rate. Note the absence of the spherical Bessel function 
in~${\cal T}_1^\up{CMB}$ in Eq.~\eqref{CMBdipole} for the observer motion.
Hence no suppression due to the spherical Bessel function for the 
observer motion.

Given that the dipole power~$C_1^\up{CMB}$ in Eq.~\eqref{CMBdipolepow}
is an integral over the dipole transfer function~${\cal T}^\up{CMB}_1(k)$
in log space with the scale-invariant power spectrum~$\Delta^2_{\cal R}(k)$,
one can read off the contributions of the individual components in 
Figure~\ref{Fig:TF} to the dipole power. Most of the dipole power
comes from the observer motion (solid), and the remaining 0.01\% of the 
dipole power 
comes from the other effects. The presence of the spherical Bessel
function~$j_1(x)$ with argument~$x=k\rbar_\star$ 
further suppresses all the contributions on small scales ($x$ larger than
the vertical lines), except
the contribution of the observer motion. The spherical Bessel functions appear
because we expanded plane waves $e^{ik\cdot x}$ 
into those with definite angular momentum, and hence no such suppression
for any contributions at the observer position $\bm{x}_{\obar}=0$.
The CMB monopole also has contributions at the observer position from the
gravitational potential \cite{BAYO21}, 
but as evident in Figure~\ref{Fig:TF} those contributions are small.
On large scales $x\ll1$ with $j_1(x)\simeq x/3$, all the contributions
add up to cancel the observer motion, such that the full dipole
transfer function~${\cal T}_1(k)$ is proportional to~$k^3$ on large scales.

Now we perform numerical computations in the standard $\Lambda$CDM 
cosmology. Given the dispersion in the relative velocity 
\beeq
\label{SIGMA}
\sigma^2_{v_\up{CMB}}:=\langle \bm{v}_\up{CMB}\cdot\bm{v}_\up{CMB}\rangle
=\int d\ln k~\Delta^2_{\cal R}(k)~k^2\TT^2_{v_\up{CMB}}~,
\eneq
the CMB dipole power is related to the relative velocity dispersion
as
\beeq
C_1^\up{CMB}={4\pi\over9}\sigma^2_{v_\up{CMB}}=4.51\times10^{-6}~,
\eneq
where $\sigma_{v_\up{CMB}}=1.8\times10^{-3}=540~\kms$.
Assuming that the relative velocity is Gaussian distributed
with the velocity dispersion~$\sigma^2_{v_\up{CMB}}$, the Maxwell-Boltzmann 
distribution can be used to derive 
the mean speed of the relative velocity and its rms deviation from the mean:
\bear
\bar v&:=&\AVE{|\bm{v}_\up{CMB}|}= \sqrt{8\over3\pi}~\sigma_{v_\up{CMB}}
=496~\kms~,\\
\sigma_{\bar v}&:=&\left\langle(v_\up{CMB}-\bar v)^2\right\rangle^{1/2}
=212~\kms~.
\enar
The Planck measurements of the CMB dipole yields the relative velocity
\cite{PLANCKdop13} 
\beeq
v_\up{CMB}=369\pm0.9~\kms~,
\eneq
along the constellations of Crater and Leo 
$ (l,~b)=(263.99^\circ\pm0.14^\circ,~48.26^\circ \pm0.03^\circ)$,
which is a bit lower than the mean relative velocity~$\bar v$, but well
within 1-$\sigma$ range. Correcting the motion of our Solar system
around the Galactic center and assuming that~$v_\ga^\up{cN}$ is constant
over the shift in the spatial positions, the relative velocity of
the CMB fluid and our Galaxy becomes
\beeq
v_\up{CMB}=627\pm22~\kms~,
\eneq
along the direction $(l,~b)=(276^\circ\pm3^\circ,~33^\circ\pm3^\circ)$,
which is again well within 1-$\sigma$ range.

\section{Dipole from Supernova Observations}
\label{DSN}
The luminosity distances from distant SNIa provides a powerful
way to probe cosmological models (see, e.g., \cite{RIFIET98,PEADET99}).
 In addition to the
redshift-distance relation in the background, the luminosity distances
also fluctuate due to the perturbations in the inhomogeneous Universe.
These fluctuations can also be used to measure the dipole 
\cite{BODUKU06,COMOET11,SARAET25},
in the same way as the CMB dipole is measured. 
The analysis of supernova light curves for the luminosity distances
is often performed by using a compilation of the supernova observations
such as the Pantheon+SH0ES data-set \cite{RIYUET22,SCBRET22,BRSCET22}.
Furthermore, measurements
of the luminosity distances in the local neighborhood are also
used to construct the ``background cosmological redshift~$z_\up{cos}$'',
which can be subtracted from the observed redshift to measure the
peculiar velocities \cite{RIPRKI95}
(see, e.g., \cite{COMOET19,SODUKU23,SODUKU24,SODUKU24b,BEALPI24}
for recent measurements).

Here we investigate what contributes to the dipole in the luminosity distance
measurements and to the peculiar velocity measurements in the local 
neighborhood.

\subsection{Luminosity distance fluctuations}

With the standardization process, measurements of individual supernovae
yield the luminosity distances to the host galaxy at the
observed redshift~$z$ along the direction~$\Nang$:
\beeq
\label{LDeq}
D_L(z,\Nang)=\bar D_L(z)\bigg[1+\de D(z,\Nang)\bigg]~,
\eneq
where the background luminosity distance is~$\bar D_L(z)=(1+z)\rbar_z$
and the dimensionless fluctuation~$\de D$ of the luminosity distance
represents the variation in the observed luminosity distance~$D_L$
(or the observed angular diameter distance~$D_A$) due to
inhomogeneities in the Universe. The fluctuation~$\de D$
in the luminosity distance can be computed by following the geodesic path
between the source and the observer and by computing the physical area
as \cite{SASAK87,BODUGA06,HUGR06,BODUKU06,CLELET12,FLDUUZ13b,BEDUET14,KAHU15a}
\beeq
\label{deD}
\de D=\dz+{\de r\over\rbar_z}-\kappa+\frac12\left(\CC^\al_\al-\CC_\para\right)~,
\eneq
where $\dz$~is again the fluctuation in the observed redshift,
$\kappa$~is the lensing convergence, $\de r$~is the distortion in the
radial position at the observed redshift, and the remaining terms represent
the gravitational potential contributions at the source position.
We follow the notation convention in \cite{YOO14a,YOSC16,YOGRET18}
(see also Appendix~\ref{details}). The lensing convergence and the radial
distortion affect the luminosity distance (or the angular
diameter distance) by changing the flux (or the physical area),
 given the observed
redshift and angle.  The full expression at the linear order
in perturbations  is composed of three distinct contributions:
contributions at the source
position, at the observer position, and along the line-of-sight direction,
each of which is essential to ensure that the full expression 
is gauge-invariant \cite{BIYO17,SCYO17}.

The fluctuation in the luminosity distance in Eq.~\eqref{deD} 
can be angular decomposed
as $\de D(z,\Nang)=\sum a_{lm}(z)Y_{lm}(\Nang)$, and the dipole 
power~$C_1^\up{LD}$ can be computed in the same way as
\beeq
\label{LDC1}
C_1^\up{LD}(z)
=4\pi\int d\ln k~\Delta^2_{\cal R}(k)~|{\cal T}_1^\up{LD}(k,z)|^2~,
\eneq
with the dipole transfer function for the luminosity distance
\begin{widetext}
\bear
\label{TFSN}
{\cal T}_1^\up{LD}(k,z)&:=&-{1\over\HH_z\rbar_z}{1\over3}k\TT_{v_m^\up{cN}}
(\eta_{\obar})+\left({1\over\HH_z\rbar_z}-2\right)
\TT_\psi(\eta_z)j_1(k\rbar_z)
+\left({1\over\HH_z\rbar_z}-1\right)k\TT_{v_m^\up{cN}}(\eta_z)j_1'(k\rbar_z) \nnn
&&
+ \int_0^{\rbar_z}d\rbar\left[{2\over\rbar_z}\TT_\psi(\eta)
-2\left(1-{1\over\HH_z\rbar_z}\right)\TT_{\psi'}(\eta)
+2\left({\rbar_z-\rbar\over \rbar_z \rbar}\right)\TT_\psi(\eta)
\right]j_1(k\rbar)~,
\enar
\end{widetext}
where we again chose the conformal Newtonian gauge for the transfer
function computation and assumed that the motion of the observer and the host
galaxies for supernova observations follows the matter motion.
Similar to the transfer function for the CMB dipole,  the transfer function
for the luminosity distance dipole consists of contributions from
the peculiar motion~$v_m^\up{cN}$ and the gravitational potential~$\psi$.
The gravitational lensing effect~$\kappa$
directly contributes to the luminosity distance dipole by
changing the observed flux of the source, but not to the CMB dipole due to
the spectral measurements.

In contrast
to the CMB dipole, the dipole in the luminosity distance
receives the contribution from the observer motion~$v_m^\up{cN}$
with a coefficient $1/\HH_z\rbar_z$, arising from the contribution in the 
radial  distortion~$\de r$, while the contributions of the observer motion
in~$\dz$ and~$\kappa$ cancel (see, e.g., \cite{YOO14a,KAHU15a,YOGRET18}).
We can then
define the {\it intrinsic velocity $\bm{v}_\up{SN}^\up{cN}$}
of the ``supernova'' fluid at the observer position today by collecting
all the contributions in the dipole transfer function except the
observer motion as
\beeq
{\cal T}_1^\up{LD}(k,z)=:{1\over\HH_z\rbar_z}\frac13k\bigg[\TT_{v_\up{SN}^\up{cN}}
(k,\eta_{\obar}) -\TT_{v_m^\up{cN}}(k,\eta_{\obar})\bigg]~,
\eneq
and hence the dipole power in the luminosity distance fluctuation
is then related to the relative velocity as
\beeq
C_1^\up{LD}(z)={4\pi\over 9}~{\sigma^2_{v_\up{LD}}\over\HH_z^2\rbar_z^2}~,
\eneq
where $\sigma^2_{v_\up{LD}}$ is defined as in Eq.~\eqref{SIGMA} with the
relative velocity $\bm{v}_\up{LD}:=\bm{v}_\up{SN}-\bm{v}_\up{o}$.
Note that only the relative velocity~$\bm{v}_\up{LD}$ is physical,
not~$\bm{v}_\up{SN}$ or~$\bm{v}_\up{o}$. While the individual components
in Eq.~\eqref{TFSN} are gauge-invariant, the dipole power only measures
the relative velocity~$\bm{v}_\up{LD}$.
Since the observed supernova fluid depends on the source redshift~$z$, we can
construct multiple supernova fluids at various redshifts and derive the
dipoles as a function of redshift. However, note that 
these dipoles of the supernova fluids at various redshifts are 
the measurements of the relative
velocity~$\bm{v}_\up{SN}$ between the observer motion and the supernova
fluids at the observer position, exactly in the same way that 
the expression for the CMB photon fluid in Eq.~\eqref{CMBdipole}
depends on the decoupling redshift~$z_\star$ but
the CMB photon fluid we measure is the photon fluid at the observer position.
Only the relative velocity at the same position
is a physical observable, not the individual velocities.

In practice, the total number of the luminosity distance measurements from
individual supernovae ($\lesssim2000$, see, e.g.,
\cite{PANTH22}) is at the moment small 
for the dipole measurements at multiple redshift slices, such that all
the luminosity distance measurements over the whole redshift range
can be collected into one sample for the dipole measurement of the full
supernova distributions, for which the dipole power is (without
the source redshift dependence in~$C_1^\up{LD}$)
\beeq
C_1^\up{LD}=4\pi\int d\ln k~\Delta^2_{\cal R}(k)~|{\cal T}_1^\up{LD}(k)|^2~,
\eneq
and the full dipole transfer function 
\beeq
{\cal T}_1^\up{LD}(k)=
\int dz~\left({d\bar N_\up{SN}\over dz}\right)~{\cal T}_1^\up{LD}(k,z)~,
\eneq
is the integral of the transfer function~${\cal T}_1^\up{LD}(k,z)$
at each redshift 
weighted by the number of individual supernovae, where the redshift
distribution $d\bar N_\up{SN}/dz$ is normalized to unity.
We assumed that the source galaxies for individual supernovae are 
uniformly distributed, but the source clustering can also contribute
to the fluctuations \cite{YOMIET19,YOO20}. In this work, we only consider
how the individual components in Eq.~\eqref{TFSN} contribute to 
the dipole power from the luminosity distance measurements.

\begin{figure}[t]
\centering
\includegraphics[width=0.5\textwidth]{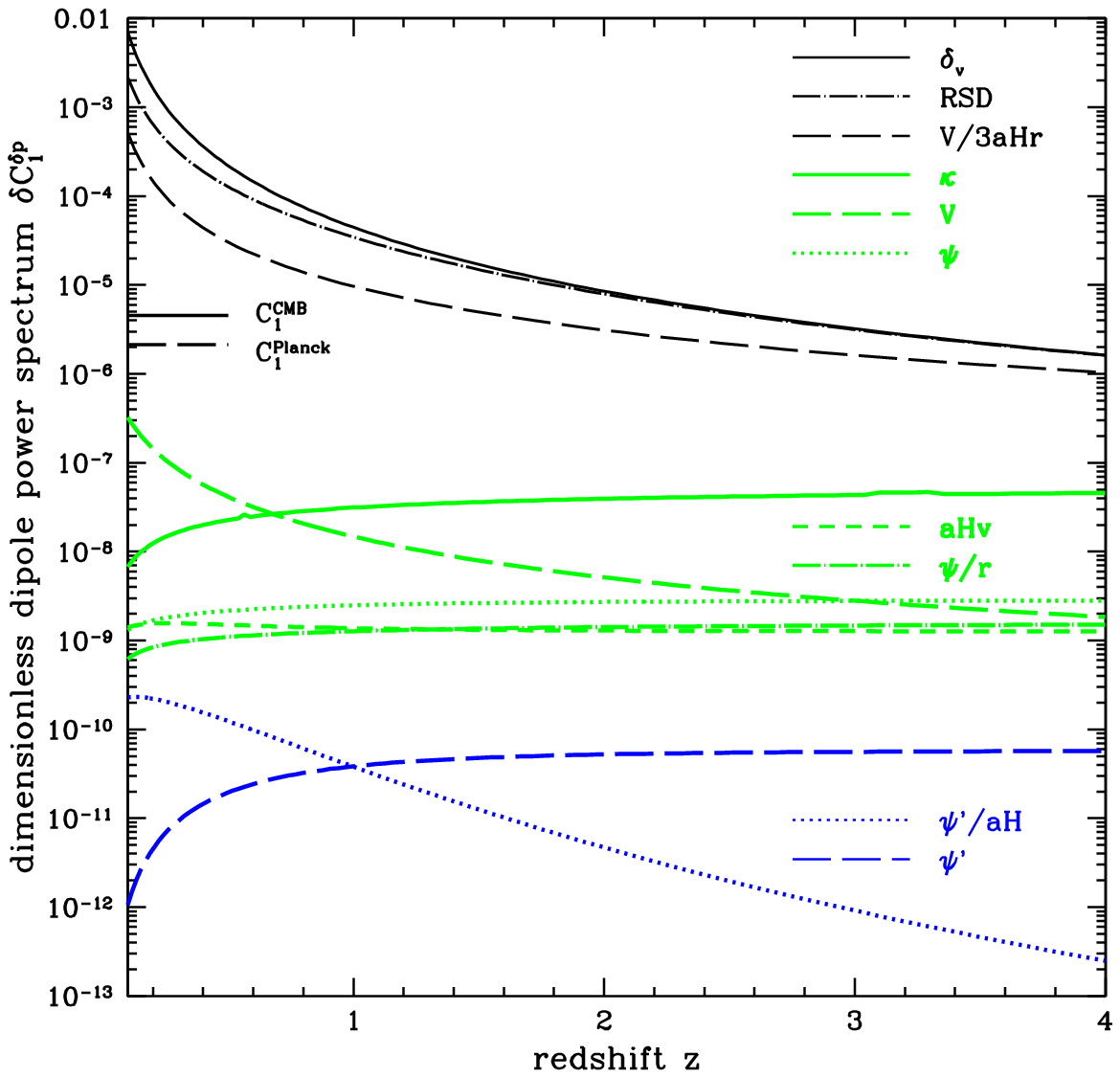}
\caption{Dimensionless dipole power spectra from the individual contributions
in Eqs.~\eqref{CMBdipole}, \eqref{TFSN}, \eqref{TFLSS}. In the dipole transfer
functions~${\cal T}_1(k)$, each transfer function $\TT_{\de p}(k,\eta_z)$
for the individual contributions~$\de p$ is multiplied by spherical Bessel 
functions, and only this product is considered for each curve to compute
the dipole power at each redshift~$z$. From the top to bottom, three
curves represent the matter 
density fluctuation~$\de_v$ (solid), the redshift-space
distortion (dot dashed), the observer motion 
$k\TT_{v_m^\up{cN}}(\eta_{\obar})/3\HH\rbar_z$
in Eq.~\eqref{TFSN} (dashed). In the middle, five green curves show
the lensing convergence~$\kappa$ (solid), the velocity $kv_m^\up{cN}$ (dashed),
the gravitational potential~$\psi$ (dotted), the velocity potential 
$\HH_zv_m^{\up{cN}}$ (short dashed), and the line-of-sight integration
of $\psi/\rbar_z$ (dot dashed). In the bottom, two blue curves show
the time derivative of gravitational potential $\psi'/\HH_z$ (dotted)
and the line-of-sight integration of $\psi'$ (dashed). The full dipole power
needs to be computed with the full transfer function, which is the sum of
all the individual components and integrated over the redshift range,
before squared and integrated in Fourier space. All these contributions
to the intrinsic velocity of the sources fall as the source redshift increases,
while the observer velocity~$\bm{v}^\up{cN}_\up{o}$
in the conformal Newtonian gauge remains 
constant. Two short horizontal lines represent the mean dipole power of CMB 
(solid: $C_1^\up{CMB}=4.5\times10^{-6}$) from our fiducial cosmological
model and the dipole power from the Planck measurements 
(dashed: $C_1^\up{Planck}= 2.1\times 10^{-6}$), which to a good approximation
represent the observer velocity~$\bm{v}^\up{cN}_\up{o}$ in the conformal
Newton gauge.}
\label{Fig:dipole}
\end{figure}

Figure~\ref{Fig:dipole} shows the individual contributions  in Eq.~\eqref{TFSN}
at a given redshift to the dipole power~$C_1^\up{LD}$. With nearly
scale-invariant power spectrum~$\Delta^2_{\cal R}(k)$, the dipole power
in Eq.~\eqref{LDC1} from
the individual components is set by the shape of the transfer functions
illustrated in Figure~\ref{Fig:TF} and the characteristic scale~$k\sim1/\rbar_z$
in the spherical Bessel functions. At redshift $z=1$, for instance, 
the comoving angular diameter distance is $\rbar_z=3406$~Mpc, and hence
the characteristic scales in~$k$ for~$j_1(x)$ and~$j_1'(x)$
at~$z_\star$ shown as short vertical
lines in Figure~\ref{Fig:TF} are accordingly shifted by the ratio of
$\rbar_\star/\rbar_z=4.07$ at $z=1$. Roughly speaking, the dipole power
in Figure~\ref{Fig:dipole} decreases with increasing redshift, simply
because the characteristic scales in~$k$ move to larger scales
(smaller wave numbers), where the transfer function decreases rapidly. 

The dominant  contribution
in the conformal Newtonian gauge to the dipole power of the
luminosity distance fluctuation arises from the observer motion (dashed).
In contrast 
to the case of the CMB, the contribution of the observer motion
has a coefficient $1/\HH_z\rbar_z$, arising from the radial distortion
$\de r/\rbar_z$. The numerical value $1/\HH_z\rbar_z$ decreases slowly
with redshift from 2.5 at $z=0.5$ to 0.7 at $z=2.5$, and its divergence
at $z=0$ is artificial due to the vanishing luminosity distance
at $z=0$. The contribution
of the observer motion in CMB without the coefficient is shown as
a short horizontal line (solid), and the redshift-dependence in~$C_1^\up{LD}$ is
entirely due to the coefficient. As pointed out \cite{SODUKU23,SODUKU24},
the dipole power in the luminosity distance fluctuation quickly vanishes
at high redshift, again due to the redshift-dependent coefficient.
One simple way to improve the signal is to weight the individual supernovae
with $\HH_z\rbar_z$, before the dipole power is computed.

The other contributions in Eq.~\eqref{TFSN} are smaller than the contribution
from the observer motion by orders of magnitude. The source velocity
(green dashed) $kv_m^\up{cN}$ at the source redshift and the 
lensing convergence~$\kappa$ (green solid) are the next leading
contributions to the dipole power. The source velocity contribution
reduces to the observer velocity (short horizontal; $C_1^\up{CMB}$)
in the limit $z\RA0$ as $j_1'(0)=1/3$.
As can be inferred in Figure~\ref{Fig:TF},
the source velocity transfer function at redshift~$z$ is between solid 
curve at $z=0$ and green solid curve at $z_\star$, but the suppression from
the spherical Bessel function~$j_1'(x)$ reduces its contribution to
 the dipole power with
increasing redshift. This contribution to the CMB dipole can also be
estimated by extrapolating the dashed curve in Figure~\ref{Fig:dipole}
to~$z_\star$, which is orders of magnitude
smaller than the contribution of the observer motion. The lensing contribution
(green solid) increases with redshift, as it accumulates the
fluctuations along the line-of-sight direction. Since the gravitational
potential changes so little over time (dashed and green dashed) 
and its transfer function is flat on large scales shown in Figure~\ref{Fig:TF},
the lensing contribution saturates at high redshift.  However, as discussed
below in detail,
the lensing contribution remains unaffected, when we consider the
redshift distribution of the sources, while the individual contributions
at the source redshift smooth out and decrease. 

Three remaining curves in the middle show the other relativistic contributions:
the gravitational potential~$\psi$ (green dotted) at the source, 
the line-of-sight integral of the gravitational potential 
$\int d\rbar~\psi/\rbar_z$ (green dot dashed),
and the velocity potential~$\HH_zv_m^\up{cN}$ (green dashed). The latter does
not contribute to~$C_1^\up{LD}$, but to~$C_1^\up{LSS}$ discussed
in Section~\ref{LSS}. While they
are small compared to the contribution of the observer motion, the
gravitational potential contribution (dotted) at~$z_\star$
is indeed the dominant contribution  
in the CMB power spectra $C_{l\neq1}^\up{CMB}$ except for the dipole.
The other remaining contributions (blue) shown in the bottom of Figure~\ref{Fig:dipole}
is even smaller,
as the gravitational potential becomes constant soon after the dark energy
becomes negligible at high redshift.
Given the forecast that the Vera~C. Rubin observatory \cite{LSST04} would
measure $N_\up{SN}\sim10^6$ supernovae, the shot-noise for the dipole power is
\beeq
C_\up{shot}={4\pi\over N_\up{SN}}\sim 10^{-5}~,
\eneq
such that only the velocity (dashed in Figure~\ref{Fig:dipole}) of the
sources at low redshift is relevant, in addition to
the observer velocity.
Note that the boost factor~$1/\HH\rbar$ enhances the velocity contributions.

\subsection{Peculiar velocity measurements}
As mentioned, the luminosity distance measurements can also be used 
as a direct measurement of our peculiar
velocity measurements in the local neighborhood
 \cite{RIPRKI95,COMOET11,SODUKU23,SODUKU24,SODUKU24b,BEALPI24}. At low redshift
($z\ll1$), the background luminosity becomes
\beeq
\bar D_L(z)=(1+z)\rbar_z\simeq{z\over H_0}~,
\eneq
and the (dimensionless) fluctuation in the luminosity distance becomes
\cite{HUGR06,BODUKU06,BEDUET14,KAHU15a,YOSC16}
\beeq
\de D\simeq V^\up{cN}_\para-{1\over\HH_z\rbar_z}\left(V^\up{cN}_\para
-V^\up{cN}_{\para,\obar}\right)
\simeq V^\up{cN}_\para-{V^\up{cN}_\para-V^\up{cN}_{\para,\obar}\over z}~,
\eneq
where we ignored all the potential contributions in Eq.~\eqref{deD},
$V^\up{cN}_\para$, $V^\up{cN}_{\para,\obar}$ 
are the line-of-sight velocities of the source
and the observer in the conformal Newtonian gauge, 
and $\HH_z\rbar_z\simeq z$ at low redshift.
Note that the velocity is defined as $\bm{v}=-\nabla v$, and the line-of-sight
velocity is $V_\para:=\hat n\cdot\bm{v}$.
Hence the observed luminosity distance in Eq.~\eqref{LDeq} at low redshift  is
\beeq
D_L(z,\Nang)\simeq{z+V^\up{cN}_{\para,\obar}-V^\up{cN}_\para\over H_0}~,
\eneq
where we used the low-redshift approximation $zV^\up{cN}_\para\simeq\OO(2)$.
By defining the ``cosmological redshift'' $z_\up{cos}:=H_0D_L(z,\Nang)$
for a given choice of the Hubble parameter~$H_0$, the peculiar velocity
of the source can be obtained (if the choice of~$H_0$ is correct) as
\beeq
\label{DZ}
\Delta z:=z-z_\up{cos}\simeq V^\up{cN}_\para-V^\up{cN}_{\para,\obar}~,
\eneq
which is in fact the relative velocity between the source and the observer
in the conformal Newtonian gauge. Note that the relative velocity is
gauge-invariant and with the low-redshift approximation the source and the
observer position are essentially the same.

Any velocity flows that shift the source and the observer together
cannot be measured, and only the relative motion can be measured, 
as apparent in Eq.~\eqref{DZ}. According to the equivalence principle,
a uniform shift in the gravitational potential or a uniform shift in
the velocity affects both the test particle and the laboratory, leaving
no observable trace. While not manifest in Eq.~\eqref{TFSN}, 
it was shown \cite{BIYO16,BIYO17,MAYO23} 
that any such fluctuations with
wavelength larger than the separation between the source and the observer
cancel each other in the luminosity distance fluctuations, 
if we used the full relativistic formula in a $\Lambda$CDM universe
(the cancellation requires not only the velocity, but also the gravitational
potential contributions).

In the local neighborhood, the peculiar velocities of the individual
sources can be measured. In particular, the supernova observations
are used \cite{WAFEHU09,WAALET23} to estimate the bulk 
velocity~$\bm{v}_\up{bulk}$. While the anomalously large amplitudes
in the measurements are odd, the presence of any
large-scale fluctuations cannot explain the deviation from the
standard $\Lambda$CDM model predictions. The relative 
velocity~$\bm{v}_\up{bulk}$ of the bulk flow can be
altered, only when such fluctuations treat the sources and the observer
differently. Even in the presence of isocurvature fluctuations,
it is difficult to provide a mechanism to discriminate the observer versus
the supernova sources (see \cite{DOMOET22,KASIJA25} for the impact of isocurvature
fluctuations on large scales).

\section{Dipole from Galaxy Surveys}
\label{LSS}
Galaxy surveys map the angular positions of distant galaxies, quasars,
or any other cosmological sources, in general as a function of its flux 
(see, e.g., \cite{YOADET00,CODAET01}). 
For galaxy clustering analysis, it is important
to have the redshift measurements of the sources. 
With the Ellis \& Baldwin test \cite{ELBA84} for the cosmological 
principle by measuring the dipole from galaxy surveys, however,
the redshift information is not essential, but a large sky coverage of 
the surveys ($f_\up{sky}\geq1/2$)
is critical for the dipole measurements, posing a distinct 
challenge in modern galaxy surveys. A few, but certainly not many galaxy
surveys meet this requirement.

The NRAO VLA Sky Survey (NVSS; \cite{COCOET98}) of radio
sources is the first large-scale survey, in which the dipole from
galaxy surveys is robustly measured
\cite{BLWA02,SINGA11,GIHU12,RUSC13,TIKOET15,TINU16}.
Subsequently, other surveys such as 2MASS \cite{SKCUET06}, 
TGSS \cite{INJAET17}, SDSS \cite{YOADET00}
have been used to measure the dipoles 
\cite{GIHU12,BEMASA18,SINGA19,TISCET24,SINGA24,SIMA24}.
Recently, by using the quasars from the WISE survey\cite{WREIET10},
 a more
precise measurement of the dipole was obtained in \cite{SEHAET21}, 
proving the general trends seen from all the previous measurements
that the dipoles from galaxy surveys are largely aligned with the 
direction of the CMB dipole, but its amplitude is larger by a factor of 
few, compared to our theoretical expectation of the Ellis \& Baldwin
formula based on the CMB dipole measurements. The discrepancy
in the dipole measurement is now significant at the 5-$\sigma$ level
 \cite{SEHAET21}. Here we discuss in detail what contributes to the
dipole from galaxy surveys.

A galaxy sample is chosen over a redshift range in a given survey
in terms of their properties such as luminosity, morphology, and color.
In particular, a constant luminosity threshold over the redshift range
(or volume-limited sample) is important to ensure the uniformity of the
chosen galaxy sample. For simplicity, we assume that galaxy samples 
are chosen in terms of a luminosity threshold~$L_t$. The observed galaxy
number counts as a function of the observed redshift~$z$ and angular
direction~$\Nang$ can then be used to derive the observed galaxy number 
density:
\beeq
\ngobs(z,\Nang)=\bar n_g(z)[1+\degi(z,\Nang)]~,
\eneq
and the number density fluctuation~$\degi$ at the linear order
in perturbations originate from two distinct effects \cite{YOO09}:
\beeq
\label{SRVO}
\degi=\de S+\de V~,
\eneq
where $\de S$ represents the contributions associated with the source
galaxy population and $\de V$ represents the contributions associated
with the observed volume.

The dominant source effect is the intrinsic galaxy fluctuation, expressed
in terms of the matter density fluctuation~$\de_v$  in the rest frame of
the source galaxies (or the proper time hypersurface \cite{YOO14b,YOO23})
and the galaxy bias factor~$b$ \cite{KAISE84}, where the subscript~$v$
indicates that the comoving gauge ($v\equiv0$) is chosen for computing
the matter density fluctuation. Two additional contributions in the source
effect arise due to our expression of the observed galaxy at the observed
redshift~$z$ and the luminosity cut of the sample 
at the observed redshift, where
the evolution bias factor and the magnification bias factor are
defined as
\beeq
e_z:={d\ln\bar n_g\over d\ln (1+z)}~,\Dquad t_L:=-2{d\ln\bar n_g\over d\ln L}~,
\eneq
and two coefficients depend on redshift and the luminosity threshold.
Our definition of the magnification bias factor is related to the slope~$s$
of the luminosity function $d\bar n_g/dL\propto L^{-s}$
or the slope~$p$ in terms of magnitude for the cumulative number density
$\log_{10}\bar n_g(\leq M)=pM+$constant as
\beeq
t_L=5p=2(s-1)~.
\eneq
Hence the source effect is
\beeq
\de S=b~\de_v-e_z~\dz_v-t_L~\de D~,
\eneq
where $\dz_v$ is the distortion~$\dz$ in the observed redshift evaluated
in the comoving gauge ($v\equiv0$) and $\de D$ is the fluctuation
in the luminosity distance in Eq.~\eqref{deD}.
If we impose more conditions for selecting galaxies in our sample
(for example, galaxy size and shape), there will be additional terms
in~$\de S$, corresponding to the fluctuation associated with the extra
conditions imposed in terms of our observables.

The volume effect~$\de V$ arises from the distortion in the relation between
the observed volume and the physical volume occupied by the observed galaxies,
and it can be readily computed by tracing the photon geodesic backward
to the source as
\beeq
\label{dV}
\de V= 3~\dz+\de g+2~{\de r\over\rbar_z}-2\kappa+H{d\over dz}\de r-\al
+V_\para~,
\eneq
where $\de g$ is the metric determinant and $V_\para$ is the line-of-sight
velocity (see Appendix~\ref{details}). In contrast to the source effect,
the volume effect is independent of how we choose our galaxy samples.
The expression in Eq.~\eqref{dV} is derived at the linear order in
perturbations, accounting for all the relativistic effects 
and it was shown \cite{YOFIZA09,YOO10,CHLE11,BODU11,JESCHI12}
that the full expression is gauge-invariant.
The dominant contributions in the volume effect are the redshift-space
distortion \cite{KAISE87} in the radial derivative of~$\de r$ and
the lensing effect~$\kappa$ \cite{NARAY89}.
The factor two of the lensing convergence~$\kappa$ in~$\de V$
and the term $t_L\de D\ni t_L\kappa$ in the source effect are collectively
referred to as the lensing magnification bias
\cite{NARAY89,BARTE95,JASCSH03,SCMEET05}.

The observed galaxy number density~$\ngobs$ at the linear order in perturbations
receives contributions from three different types, exactly in the same
way as the CMB temperature anisotropies: one from the source position,
one from the observer position, and one along the line-of-sight direction.
Once we decompose $\deobs(z,\Vang)=\sum a_{lm}(z)Y_{lm}(\Vang)$,
the dipole transfer function can be straightforwardly computed as
\begin{widetext}
\bear
\label{TFLSS}
{\cal T}^\up{LSS}_1(k,z)&:=&\left(1-h_z+{t_L\over\HH_z\rbar_z}\right)
{k\over3}\TT_{v_m^\up{cN}}  (\eta_{\obar})
  +\left[h_z-3+t_L\left(1-{1\over\HH_z\rbar_z}\right)
\right]k\TT_{v_m^\up{cN}}(\rbar_z)j_1'(k\rbar_z)
+{k^2\over\HH_z}\TT_{v_m^\up{cN}}(\rbar_z)j_1''(k\rbar_z)  ~~~~~~~~~~~\\
&&
+\left\{b~\TT_{\de_v}(\rbar_z)-e_z\HH_z\TT_{v_m^\up{cN}}(\rbar_z)
+\left[h_z-4+t_L\left(2-{1\over\HH_z\rbar_z}\right)\right]
\TT_\psi(\rbar_z)+{1\over\HH_z}\TT_{\psi'}(\rbar_z)
\right\}j_1(k\rbar_z) \nnn
&&+ 
\int_0^{\rbar_z}d\rbar \left\{\left({4-2t_L\over\rbar_z}\right)\TT_\psi(\rbar)+
\left[2\left(h_z-3\right)+2t_L\left(1-{1\over\HH_z\rbar_z}\right)\right]
\TT_{\psi'}(\rbar)
-2(t_L-2)\left({\rbar_z-\rbar\over \rbar_z \rbar}\right)\TT_\psi(\rbar)
\right\}j_1(k\rbar) ~, \nn
\enar
where we again chose the conformal Newtonian gauge and
suppressed the $k$-dependence in the transfer functions~$\TT_{\de p}$,
and the redshift-dependent coefficient~$h_z$ is defined
\cite{SCYOBI18,GRSCET20} as
\bear
h_z:=e_z+{2\over\HH_z\rbar_z}+{\HH'_z\over\HH^2_z}~,\Tquad
{\HH'\over\HH^2}=1+{\dot H\over H^2}=-\frac12\sum_i(1+3w_i)\Omega_i
=-\frac32\sum_i(1+w_i)\Omega_i-\Omega_k~,
\enar
\end{widetext}
with the relation $1=\Omega_k+\sum\Omega_i$ and $w_i$ being the equation
of state for individual component.
The first contribution $k\TT_{v_m^\up{cN}}/3$ is again the contribution
of our peculiar motion at the observer position. 
 Evident in Eq.~\eqref{TFLSS}, the dipole moment in galaxy surveys 
{\it not only} arises from our peculiar motion, {\it but also}
receives contributions from other sources such as the matter density 
fluctuation~$\TT_{\de_v}$, the redshift-space distortion
$k^2\TT_{v_m^\up{cN}}/\HH_z$, 
the gravitational lensing~$(\rbar_z-\rbar)\TT_\psi/\rbar_z\rbar$, and other
velocity and relativistic contributions, consisting of three distinct
types of contributions, 
exactly in the same way for CMB in Eq.~\eqref{CMBdipole} and
the luminosity distance fluctuation in Eq.~\eqref{TFSN}.

In contrast
to measuring the three-dimensional clustering of galaxies
with the observed galaxy number density $\ngobs(z,\Nang)$,
the dipole measurements in galaxy surveys are performed by collecting all
the galaxies in a given redshift range (or projecting along the line-of-sight
directions) to construct galaxy sky maps~$\ngtwo(\Nang)$. The number of
galaxies in a unit solid angle is $dN_g(\Nang)=\ngtwo(\Nang)d\Omega$,
and the angular number density of galaxies is then related to~$\ngobs$ as
\beeq
\label{map}
\ngtwo(\Nang)=\int dz~\left({d\bar V\over dz~d\Omega}\right)\ngobs(z,\Nang)~,
\eneq
where $(d\bar V/dzd\Omega)=\rbar^2_z/H_z(1+z)^3$ is  the physical volume
element in a unit redshift~$dz$ and a unit solid angle~$d\Omega$ 
in a background universe and the integration is over the full redshift 
range of the sample. In the limit the redshift range 
goes to zero ($\Delta z\RA0$), the angular number
density reduces to the observed galaxy number density at a given redshift
$\ngtwo(\Nang)\propto\ngobs(z,\Nang)$, but the physical volume factor 
at the same time becomes zero, giving rise to an infinite shot-noise 
contribution (see Appendix~\ref{shotnoise}).
Hence the redshift depth should be chosen sufficiently large
enough to ensure a reasonable signal-to-noise ratio.

The galaxy sky maps in Eq.~\eqref{map} can be angular decomposed 
in the same way as
\beeq
{\ngtwo(\Nang)\over \bar\Sigma_g}=\sum_{l\geq1} a_{lm}Y_{lm}(\Nang)~,
\eneq
after removing the background average (angular) number density 
\beeq
\bar\Sigma_g:=\int dz~\left({d\bar V\over dz~d\Omega}\right)\bar n_g(z)~.
\eneq
The dipole power~$C_1^\up{LSS}$ can then be again computed as
\beeq
C_1^\up{LSS}=4\pi \int d\ln k~\Delta_{\cal R}^2(k)\left|{\cal T}_1^\up{LSS}
(k)\right|^2~,
\eneq
where we defined the full dipole transfer function
\beeq
\label{LSSTF}
{\cal T}_1^\up{LSS}(k):={1\over\bar\Sigma_g}
\int dz~\left({d\bar V\over dz~d\Omega}\right)
\bar n_g(z)~{\cal T}_1^\up{LSS}(k,z)~.
\eneq
While the dipole in the galaxy sky maps is 
much more complicated than in Eq.~\eqref{CMBdipole} of CMB temperature
anisotropies or in Eq.~\eqref{TFSN} of the luminosity distance fluctuations,
we can define in the same way the {\it intrinsic velocity} 
$\bm{v}_g^\up{cN}$ of the galaxy fluid at the observer position today
by adding all the contributions in the dipole transfer function
${\cal T}_1^\up{LSS}(k)$ except the observer motion as
\beeq
\label{intLSS}
{\cal T}_1^\up{LSS}(k)=:{\cal M}~\frac13k\bigg[\TT_{v_g^\up{cN}}
(k,\eta_{\obar}) -\TT_{v_m^\up{cN}}(k,\eta_{\obar})\bigg],
\eneq
with the coefficient
\beeq
{\cal M}:={1\over\bar\Sigma_g}\int dz~\left({d\bar V\over dz~d\Omega}\right)
\bar n_g(z)\left(h_z-1-{t_L\over\HH_z\rbar_z}\right)~,
\eneq
and hence the dipole power from galaxy surveys is a measure of
the relative velocity as
\beeq
C_1^\up{LSS}={4\pi\over9}{\cal M}^2\sigma^2_\up{LSS}~.
\eneq

For the moment, {\it if} we ignore all the contributions in~$\ngobs$
except one due to the peculiar motion of the observer (i.e., 
we ignore the {\it intrinsic velocity} contribution or the kinetic dipole),
the observed angular  density of galaxies is then
\beeq
\ngtwo(\Nang)\approx \bar\Sigma_g\left[1-{\cal M}(\pa_\para v)_{\obar}\right]~,
\eneq
and note that the line-of-sight velocity of the observer in the conformal
Newtonian gauge is $V_\para=\bm{V}_{\obar}\cdot\Nang=-(\pa_\para v)_{\obar}$.
Given the angular dependence of the dipole, the coefficient~$\cal{M}$
 can be measured from large-scale
surveys, and it was shown \cite{NADUET21} that
this coefficient~$\cal M$ corresponds to the dipole amplitude
of the Ellis-Baldwin test in the absence of redshift evolution of
(see also \cite{DABO22,GUPIET23}). In fact, \cite{HAUSE24} showed
that the correspondence is generally valid even with redshift evolution.

Figure~\ref{Fig:dipole} shows the individual contributions in Eq.~\eqref{TFLSS}
at the source redshift to the dipole power~$C_1^\up{LSS}$. Each contribution
is again computed at a given redshift, and only one contribution is 
considered in computing the dipole power in Figure~\ref{Fig:dipole}. 
Compared to the CMB and the luminosity distance fluctuations, the notable 
difference in the observed galaxy number density fluctuation in 
Eq.~\eqref{TFLSS} is the existence  of the matter density fluctuation~$\de_v$
with the galaxy bias factor~$b$. On large scales, galaxies are biased
tracers of the underlying matter distribution \cite{KAISE84}, and
the transfer function for the matter density contribution at $z=0$
(dot dashed) in the comoving gauge is larger than any other contributions 
in Figure~\ref{Fig:TF} by orders of magnitude. Since the transfer function
scales as $\TT_{\de_v}\propto k^2$ on large scales and $\TT_{\de_v}\propto k^{1/2}$
on small scales, its contribution to the dipole power (solid) in
Figure~\ref{Fig:dipole} decreases with increasing redshift, again due to
the suppression from the spherical Bessel function~$j_1(x)$, in addition
to the decreasing growth factor~$D(z)$. Note that the linear bias factor is
not considered in Figure~\ref{Fig:dipole} and it is expected 
to grow $b(z)-1=(b_0-1)/D(z)$ in a simple model \cite{FRY96}, 
where the growth factor is normalized to unity at $z=0$.

As discussed in Section~\ref{subtle},
the dipole measurements probe the fluctuations in the observed galaxy
samples or CMB temperature anisotropies at the characteristic
scale $k\sim1/\rbar_z$, and the fluctuations from the matter density
or the observer motion are never zero at any redshift.
At high redshift, the contribution from the observer motion
($C_1^\up{CMB}$: short horizontal line in Figure~\ref{Fig:dipole})
dominates over the other contributions, while at low redshift 
the contribution from the matter density fluctuation
is significantly larger than the observer motion 
\cite{BLWA02,GIHU12,RUSC13,YOHU15,NUTI15,TINU16,NADUET21,DABO22,DOMOET22},
even after accounting for the projection over the redshift range (see below).
The dipole contribution from the local
structure (density fluctuation) is also referred to as local
dipole or clustering dipole.

The other interesting contribution to~$C_1^\up{LSS}$ is the redshift-space
distortion  \cite{KAISE87} (dot dashed) in Figure~\ref{Fig:dipole},
which arises from the volume effect with
the line-of-sight velocity contribution to the observed
galaxy number density fluctuation. Being a spatial derivative of the 
line-of-sight velocity, the redshift-space distortion is as important
as the matter density fluctuation to the three-dimensional clustering of
galaxies. However, once the galaxy sky map~$\ngtwo$ is constructed by
collecting all the galaxies along the line-of-sight, the redshift-space
distortion contribution to the dipole power is significantly reduced,
as the volume effect contributes to the observed galaxy sky map
only around the boundary of the redshift range.

Two remaining contributions in Figure~\ref{Fig:dipole}
that are not yet discussed
in~$C_1^\up{CMB}$ or~$C_1^\up{SN}$ are the velocity potential~$\HH_zv_m^\up{cN}$
(green short dashed) and the time derivative of the gravitational
potential $\psi'/\HH_z$ (blue dotted) in Figure~\ref{Fig:dipole}.
The dimensionless velocity potential is $\HH_zv_m^\up{cN}=1-1/\Sigma(z)$
with little time-dependence \cite{YOGO16} in a $\Lambda$CDM universe,
where $\Sigma(z):=1+3\Omega_m(z)/2f$ increases from 1.9 at $z=0$
to 2.5 in the matter dominated era. The time-derivative of the potential
is $\psi'/\HH_z=\psi(f-1)$, and it vanishes once $f\RA1$ in the matter
dominated era. Both effects are negligible contributions to $C_1^\up{LSS}$.

As mentioned, the individual contributions to the dipole power from
galaxy surveys in Figure~\ref{Fig:dipole}
are illustrative, but not directly observable, due to the infinite
shot-noise contributions, arising from the zero volume in a redshift
slice with its depth $\Delta z\RA0$ (see Appendix~\ref{shotnoise}).
In real surveys, the source samples for the dipole measurements
 are constructed by collecting all the galaxies at all redshifts
that is brighter than some flux threshold, and it is quite often the case
there exist no redshift information for individual galaxies in the sample.
Here we numerically compute the contributions of the intrinsic velocity
to the dipole power by assuming various redshift distributions of the source
galaxies.

\begin{figure}[t]
\centering
\includegraphics[width=0.5\textwidth]{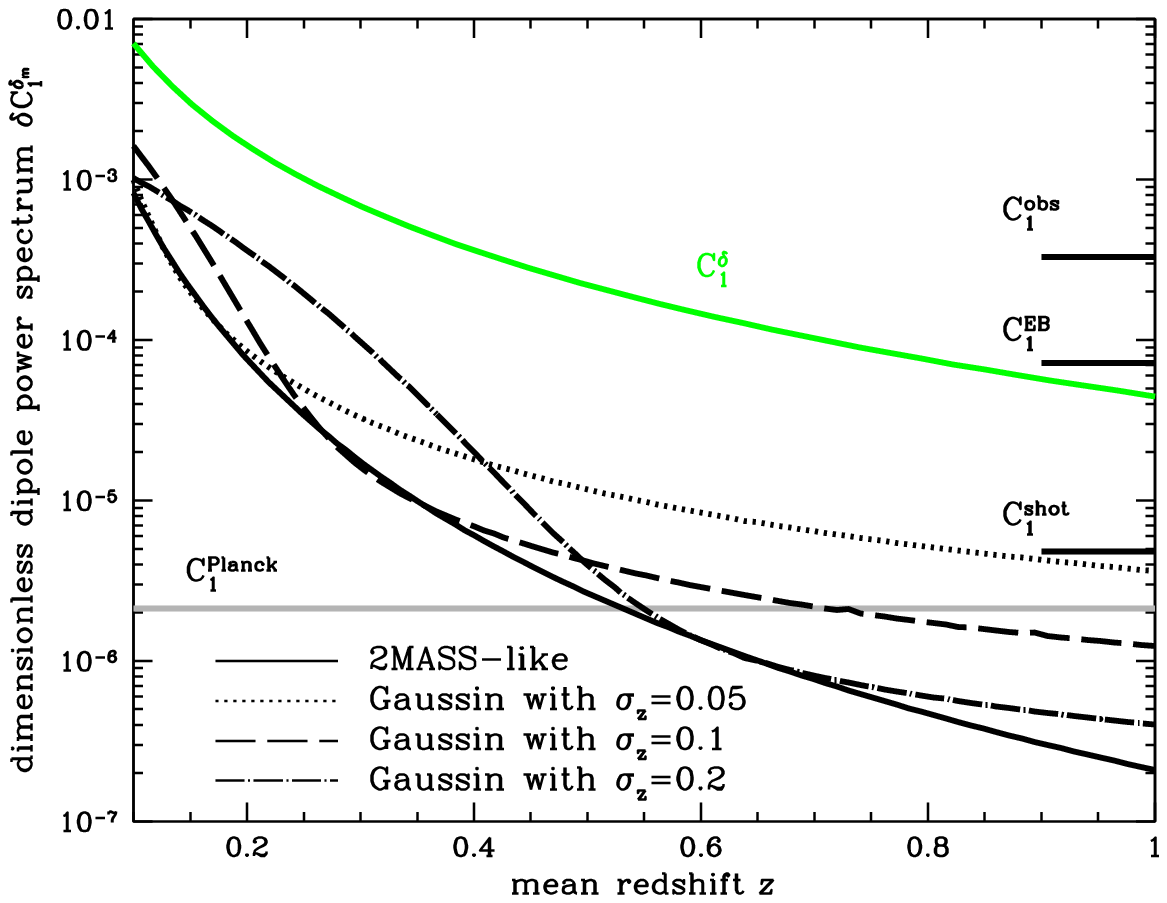}
\caption{Dimensionless dipole power spectra from the matter density
fluctuations integrated over various redshift distributions.
Solid curve shows the dipole power with the distribution in Eq.~\eqref{2MASS}
as a function of the mean redshift~$z$. The redshift distribution
of the CatWise sample is similar to the solid curve with mean redshift~$z=1$.
Dotted, dashed, and dot-dashed curves show the dipole power with Gaussian
redshift distribution in Eq.~\eqref{GAUSS}, and the variance for the
Gaussian distributions is set $\sigma_z=0.05$, 0.1, and 0.2, respectively.
In comparison, the green solid curve shows the dipole power from the
matter density fluctuations~$C_1^\de$ at each redshift slice shown in 
Figure~\ref{Fig:dipole}. The density contributions are greatly reduced
in amplitude, once averaged over the redshift distributions. Three
horizontal lines at~$z=1$ indicate the observed dipole power $C_1^\up{obs}$
from the CatWise survey, the dipole power $C_1^\up{EB}$ expected from
the Ellis-Baldwin formula for the survey, and the shot-noise power.
The dipole power from the Planck measurements is shown as the gray
horizontal line. Note that the matter density contributions in this plot
should be
multiplied by the bias factor~$b\approx2$ and the Ellis-Baldwin 
coefficient~${\cal M}\approx 6$, and hence the dipole power~$C_1\propto
b^2{\cal M}^2$ is boosted by $\approx140$ for various curves to be
compared to the observed dipole power~$C_1^\up{obs}$.}
\label{VALUE}
\end{figure}

Among all the individual contributions to the intrinsic velocity of the
sources in Eq.~\eqref{TFLSS}, we only consider the contribution of the
matter density fluctuation. As evident in Figure~\ref{Fig:dipole},
the density fluctuation (solid)
is the dominant contribution to the dipole power.
While the redshift-space distribution (dot-dashed) is comparable to the
density fluctuation, their impact is greatly reduced in galaxy samples
obtained by integrating over all redshift range, rather than by collecting
galaxies in small redshift bins, as the effect originates from the
radial distortion in a narrow redshift bin \cite{CHLE11}.
Arising from the fluctuation along the line-of-sight direction, the contribution
of the gravitational lensing (green solid) remains similar in
both the redshift-bin and projected samples. Their contribution is, however,
several orders of magnitude below the density fluctuations at $z\leq2$,
though its contribution indeed dominates at higher redshift.

Figure~\ref{VALUE} shows the dipole power from the matter density fluctuations
of the sources distributed over various redshift range. The normalized
redshift distribution of the galaxy samples is defined as
\beeq
{dN_g\over dz}:={4\pi f_\up{sky}\over\bar\Sigma_g}
\left({d\bar V\over dz~d\Omega}\right)\bar n_g(z)~,
\eneq
and we consider two redshift distributions: A simple Gaussian
distribution in terms of the mean redshift~$\bar z$ and its 
variance~$\sigma_z$:
\beeq
\left({dN_g\over dz}\right)_\up{G}={1\over\sqrt{2\pi}\sigma_z}~
\exp\left[-{(z-\bar z)^2\over 2\sigma_z^2}\right]~,
\label{GAUSS}
\eneq
and a more realistic distribution that is used to model the 2MASS
observations \cite{GIHU12}:
\beeq
\left({dN_g\over dz}\right)_\up{R}={3z^2\over2(\bar z/1.412)^3}~\exp
\left[-\left({z\over \bar z/1.412}\right)^{3/2}\right]~,
\label{2MASS}
\eneq
where the mean redshift ($1.065 \bar z$) and the peak ($0.858\bar z$)
of the distribution are given in terms of the parameter~$\bar z$.
The former describes the redshift-bin samples, and the latter has the
typical power-law distribution combined with an exponential cut-off at
high redshift. The CatWise sample \cite{EIMAET20} is well described by the
latter with $\bar z\approx 1$ \cite{SEHAET21,DABO22}.

The contribution from the matter density fluctuations (green solid curve)
in Figure~\ref{VALUE} is much larger than the dipole power~$C_1^\up{Planck}$
from the observer motion (horizontal gray line; redshift-independent).
However, the spherical Bessel function~$j_1(k\rbar_z)$ is averaged 
over a range of redshift with the weight given by $(dN_g/dz)$
as in Eq.~\eqref{LSSTF}, leading
to the reduction in amplitude at a given~$k$ and hence the lower dipole 
power~$C_1$, compared to the dipole power at each redshift slice (green
solid curve). The suppression scale~$k$ is set by the condition
$k\Delta \rbar_z\approx$ few, such that the integration over the redshift
range covers a few periods of the oscillations.
This argument is indeed borne out by various curves in 
Figure~\ref{VALUE}. Three curves (dotted, dashed, and dot-dashed)
with the Gaussian redshift distribution show that the dipole power from
the matter density fluctuations at a given redshift is reduced by
a factor ten (dotted with $\sigma_z=0.05$) or two orders-of-magnitude 
(dot-dashed with $\sigma_z=0.2$) in amplitude. At lower redshift 
$\bar z\simeq\sigma_z$, where $k\rbar_z\approx k\Delta\rbar_z$,
however, the average over the redshift range probes the first peak of
the oscillations, and hence it adds up, rather than averages out to zero.
A similar trend is also shown in the dark solid curve for the 2MASS
distribution, which resembles the redshift-bin sample at lower values
of~$\bar z$
but a broader distribution at higher values of~$\bar z$.
In \cite{NADUET21}, the same conclusion was drawn by using a similar
but different parametrization of the redshift distribution.

Analyzing the CatWise samples, it was found \cite{SEHAET21} that
the observed dipole power~$C_1^\up{obs}$ is a factor four higher (a factor
two in terms of~$\sigma_\up{LSS}$) than expected dipole power~$C_1^\up{EB}$
from the Ellis-Baldwin prediction \cite{SEHAET21,DOMOET22}
based on two parameters ($x\approx1.7$
and~$\alpha\approx1.3$, i.e., ${\cal M}\approx6$) and
the Planck dipole power~$C_1^\up{Planck}$: 
\beeq
C_1^\up{EB}:={\cal M}^2C_1^\up{Planck}~.
\eneq
The dipole power $C_1^\de\simeq2\cdot10^{-7}$ 
from the matter density fluctuations at~$\bar z=1$
needs to be further multiplied
 by the bias factor~$b(z)$ of the samples and the Ellis-Baldwin
coefficient~${\cal M}$, which leads to an approximate factor 140 in~$C_1^\de$.
Nevertheless, this contribution is still smaller by a factor ten 
around~$\bar z=1$, not enough to explain the observed dipole power.
The nonlinearity in the matter density fluctuation can significantly
boost the contribution (see, e.g., \cite{SMPEET03}), 
but only at redshift $\bar z\leq0.5$,
at which the peak contribution of the matter density fluctuation
is at $k\approx1/\rbar_z\approx 7\cdot10^{-4}~\hmpci$.

The shot-noise contribution to the dipole power~$C_1^\up{shot}$ also
arises, as galaxies are discrete objects, which is not aligned with
the observer motion. This contribution studied \cite{NADUET21}
in detail is not boosted by the Ellis-Baldwin coefficient~${\cal M}$,
and it is too small to make up the difference in observations.
In conclusion, the matter density contribution to the intrinsic dipole power
becomes quickly negligible at higher redshift,
 as it samples the matter power spectrum at
$k\simeq1/\rbar_z$ and it averages out with larger width in the redshift
distribution. However, it is important to keep in mind
that the intrinsic dipole mostly
from the matter density fluctuations can be as large as the contribution
of the observer motion at $z\simeq0.5$, where various curves cross
the gray horizon line~$C_1^\up{Planck}$. 

Similar calculations have been performed 
\cite{GIHU12,SEHAET21,NADUET21,DABO22,DOMOET22} in the past,
regarding the contributions to the dipole power from the
matter density fluctuations with various redshift ranges of the galaxy 
samples.

\section{Discussion and Summary}
\label{summary}
We have critically investigated the origin of the dipole signals 
from the cosmic microwave background anisotropies (CMB), 
supernova observations, and 
galaxy surveys. The observed sources in these large-scale surveys
such as CMB photons, supernovae, 
 or galaxies collectively form an effective fluid, 
in which an intrinsic velocity of the sources can be assigned  
at the observer position. The dipole signals from these large-scale
surveys arise from the relative velocity between the observer velocity
and the intrinsic velocity of the sources. While the observer velocity is
common in all dipole signals, the intrinsic velocities differ for each
large-scale structure probe. However, the contributions to the intrinsic
velocities fall steeply as the effective redshift of the sources increases,
and the dominant contribution to the dipole signals is the observer velocity.

This interpretation of the dipole signals, however, depends on our choice
of gauge condition, which is implicitly made in the literature 
to be the conformal Newtonian gauge. 
In general relativity, only the relative velocity between two motions
at the same spacetime position is physically meaningful, not the absolute 
velocity of one motion, or the relative motion at two different positions.
Measurements of the cosmic dipoles are indeed
the measurements of the relative velocities between the source and the
observer velocities, but the observer velocity in the conformal
Newtonian gauge is to a good approximation much larger than
the source velocity in the same gauge. Since these
relative velocity measurements
depend on the observer position, there is no common
rest frame in which observers at any spacetime position see no dipoles.

Beside the observer motion, the next-leading
 contribution to the dipole power arises from the matter
density fluctuations in the comoving gauge. As evident 
in Figure~\ref{Fig:dipole}, its contribution is dominant, but its effective
contribution is greatly reduced in amplitude upon averaging over the
redshift distributions of the samples, as shown in Figure~\ref{VALUE}.
In practice, the
intrinsic velocity from the matter density fluctuations is
negligible for the samples, as long as effective redshift distribution
is larger than $z=0.5$.
For example, the observed dipole power~$C_1^\up{obs}$ 
from the CatWise survey is larger by a factor 30 than~$C_1$ from the
matter density fluctuations, and of course twice larger than the
Ellis-Baldwin prediction~$C_1^\up{EB}$ from the observer motion measured
in the Planck observations \cite{SEHAET21,DOMOET22}.
This observational tension detected
in many galaxy/quasar samples with various statistical significance 
\cite{GIHU12,BEMASA18,SINGA19,SEHAET21,TISCET24,SINGA24}
constitutes a potential challenge to the standard model of cosmology.

In our numerical computation, we have assumed the standard $\Lambda$CDM
model, but our equations are valid for beyond-the-standard models such as
isocurvature models under the assumption that there is no significant
anisotropic pressure. Furthermore, 
we have assumed that the observer moves together with
the matter component and only the linear-order contributions were
accounted for. Of course, the observer is bound in the Earth, which moves
around the Sun and the Galactic center. Such motions are indeed corrected
for in the observations, and numerical simulations show that dark matter
halos that can host galaxies like ours move as a matter component, or
no velocity bias \cite{CHZHET18}.
 Nonlinearities in the matter density fluctuations
can change the dipole signals, in particular for galaxy surveys
(no density contributions in CMB or SN surveys),
though the scales $k\sim1/\rbar_z$ we probe for the dipole measurements
are often in the linear regime ($k\ll 0.1~\hmpci$), where the nonlinear
boost in the matter density fluctuation is negligible. 
Unless the effective
redshift of the galaxy samples is less than $z\lesssim0.5$, our calculations
based on the linear-order perturbation theory remain robust, and the
intrinsic velocity contributions are small.\\

We acknowledge useful discussions with  Ruth Durrer,
Sebastian von Hausegger, and Subir Sarkar.
This work is support by the Swiss National Science Foundation
and a Consolidator Grant of the European Research Council.

\newcommand{\dD}{\de\DDD}
\newcommand{\RR}{{\cal R}}
\newcommand{\bobs}{\bar {\rm o}}
\newcommand{\etc}{\noindent $\bullet$ }
\newcommand{\Kang}{\hat{\bm{k}}}
\newcommand{\be}{\beta}
\newcommand{\Si}{\Sigma}
\newcommand{\drr}{\delta r}    
\newcommand{\VV}{\mathcal{U}}  
\newcommand{\dg}{\delta g}      
\newcommand{\ax}{\alpha_{\chi}}  
\newcommand{\px}{\varphi_{\chi}} 
\newcommand{\LL}{\mathcal{L}}

\appendix

\section{Detailed Calculations}
\label{details}

\subsection{Notation Convention}
Here we present our notation convention used in the text.
The metric perturbations around the Robertson-Walker metric are defined
in terms of four variables $\al$, $\be$, $\varphi$, $\ga$ as
\bear
&&
\delta g_{00}=-2~a^2\al~,\Tquad
\delta g_{0\alpha}=-a^2\nabla_\al\be~,~~~~~\\
&&
\delta g_{\alpha\beta}=2~a^2\left(\varphi~\de_{\al\be}+\nabla_\al\nabla_\be\ga
\right)~,
\enar
where we only consider the scalar perturbations. 
Fluid velocities including the observer motion are described by
four velocity vectors~$u^\mu$ subject to the time-like condition
$-1=u_\mu u^\mu$, and the four velocity vector is parametrized by
\beeq
u^\mu={1\over a}\left(1-\al,~-\nabla^\al U\right)~,
\eneq
where we introduced the scalar perturbation~$U$ for the spatial velocity
component. The indicies~$\mu,\nu$ represent the spacetime components,
while the indicies~$\al,\be$ represent the spatial components.

These perturbations are gauge-dependent, or they change their values,
depending on our choice of gauge condition. Hence, it is important to ensure
that the our theoretical description of physical observables is 
gauge-invariant. We presented in the text our
gauge-invariant expressions for the cosmological observables without choosing
any gauge conditions, but in general
our computation is performed in the conformal Newtonian gauge,
in which the spatial gauge is fixed with $\ga\equiv0$ and the temporal gauge
is fixed with $\be\equiv0$, so that
the combination $\chi:=a(\be+\ga')$
vanishes. Two gravitational potentials~$\psi,\phi$
in the conformal Newtonian gauge
correspond to the gauge-invariant Bardeen variables \cite{BARDE80}:
\beeq
\psi=\al-\frac1a\chi'~,\Tquad \phi=\varphi-H\chi~.
\eneq
The Newtonian gauge velocity corresponds to another
gauge-invariant Bardeen variable \cite{BARDE80}
\beeq
v^\up{cN}=v-{1\over a}\chi~,
\eneq
where we defined a useful combination $v:=U+\beta$.
The matter density fluctuation~$\de$ in general is gauge-dependent, but
the density fluctuation in the rest frame of matter is also
gauge-invariant,
which corresponds to the comoving-gauge density fluctuation:
\beeq
\de_v=\de+3\HH v~,
\eneq
where $\HH=aH$ is the conformal Hubble parameter.

\subsection{Observed Galaxy Number Density and Luminosity Distance}
The galaxy number density~$n_g(x^\mu)$ at a given position~$x^\mu$ can be
split into the background number density~$\bar n_g(\eta)$ and the
fluctuation~$\de$ around the mean: $n_g(x^\mu)=\bar n_g(\eta)(1+\de)$.
In the simplest model of galaxy formation \cite{PRSC74,KAISE84},
the galaxy number density fluctuation is related to the matter density
fluctuation with a linear bias factor~$b$ in the rest frame of galaxies
and matter, which coincides with the comoving gauge ($v\equiv0$)
at the linear order in perturbations. Hence, we obtain
the expression for the galaxy number density as
\beeq
n_g(x^\mu)=\bar n_g(\eta_v)(1+b~\de_v)~,
\eneq
where the subscript~$v$ indicates that the quantities are evaluated in
the comoving gauge. When expressed at the observed redshift~$z$, instead 
of a time coordinate~$\eta$, the galaxy number density becomes
\beeq
n_g=\bar n_g(z)(1+b~\de_v-e_z~\dz_v)~,
\eneq
where the distortion~$\dz$ in the observed redshift at the 
position~$\eta(\bar z)$ is defined as
\beeq
1+z=:(1+\bar z)(1+\dz)~,
\eneq
and the evolution bias factor is defined as
\beeq
e_z:={d\ln\bar n_g\over d\ln(1+z)}~.
\eneq
Note that the redshift~$\bar z$ is just an expression for the time 
coordinate~$\eta$, and $e_z=3$ corresponds to a constant comoving number
density such as the matter component. Both quantities~$\de_v$ and~$\dz_v$
are gauge invariant:
\beeq
\dz_v:=\dz+\HH v~.
\eneq

Furthermore, the galaxy sample is often selected with a threshold 
luminosity~$L_t$, i.e., $n_g=n_g(L\geq L_t)$,
where the threshold luminosity at the observed position is defined
in terms of the threshold flux~$f_t$ and the background luminosity 
distance~$\bar D_L(z)$ as
\beeq
L_t:=4\pi\bar D_L^2(z)f_t~.
\eneq
In the presence of inhomogeneities in the Universe, the luminosity
distance~$D_L(z,\Nang)$ at the observed position fluctuates around
the background luminosity distance~$\bar D_L(z)$ as
\beeq
D_L(z,\Nang)=\bar D_L(z)(1+\dD)~.
\eneq
The (dimensionless) fluctuation~$\dD$ in the luminosity distance
is the same as the fluctuation in the angular diameter distance,
and its gauge-invariant expression can be obtained 
\cite{SASAK87,BODUGA06,HUGR06,BODUKU06,CLELET12,FLDUUZ13b,BEDUET14,KAHU15a}
by computing
the physical area subtended by the observed angle $d^2\hat n$ at
the observed redshift~$z$ as
\beeq
\dD=\dz+{\drr\over\rbar_z}-\kappa+\frac12\left(\CC^\al_\al-\CC_\para\right)~,
\eneq
where $\drr$ is the radial distortion, $\kappa$ is the convergence,
and the bracket at the source position is 
\beeq
 \CC^\al_\al-\CC_\para=2\varphi+\Delta\ga
-\pa_\para^2\ga~.
\eneq
The luminosity distance observation in supernova surveys yields~$D_L(z,\Nang)$,
including not only the background luminosity distance~$\bar D_L(z)$,
but also the fluctuation~$\dD(z,\Nang)$.

Hence, the physical luminosity~$L_p$ set by the threshold flux~$f_t$
at the position of galaxies is different from~$L_t$:
\beeq
L_p=4\pi D_L^2(z)f_t= L_t(1+2~\dD)~,
\eneq
and the galaxy sample with $f\ge f_t$ at the observed redshift 
corresponds to 
\beeq
\bar n_g(L\geq L_p)=\bar n_g(L_t)(1-t_L~\dD)~,
\eneq
where $\bar n_g(L_t):=\bar n_g(L\geq L_t)$ for simplicity
and we defined the coefficient~$t_L$ evaluated at~$L_t$ as
\beeq
t_L:=-2~{d\ln\bar n_g(L)\over d\ln L}~.
\eneq
If the luminosity function is approximated as a power-law
~$d\bar n_g/dL\propto L^{-s}$ at the threshold
or the cumulative number density~$\bar n_g(L)\propto L^{-s+1}$, 
the coefficient~$t_L$ is related to the power-law slope~$s$ as
\beeq
t_L=2(s-1)=:5p~,\Tquad p=0.4(s-1)~,
\eneq
where  $p$ is the slope in terms of magnitude $M=-2.5\log_{10}L+$const,
or 
\beeq
\log_{10}\bar n_g=-2.5p\log_{10}L+
\up{const.}=p~M+\up{const.}
\eneq
In the observed galaxy number density, there exists additional
contribution $2~\dD$ from the volume fluctuation~$\de V$ described in 
Eq.~\eqref{SRVO},
such that the total contribution is
\beeq
-t_L\dD+2~\dD\simeq(t_L-2)\kappa=(2s-4)\kappa=(5p-2)\kappa~,
\eneq
and the coefficient of~$\kappa$ is known as the magnification bias factor
\cite{NARAY89,BARTE95,JASCSH03,SCMEET05},
where we used $\dD\simeq-\kappa$~. Hence, the source effect~$\de S$
of the observed galaxy fluctuation is composed of 
\beeq
\de S=b~\de_v-e_z~\dz_v-t_L~\dD~.
\eneq

The observed galaxy fluctuation also arises from the volume effect~$\de V$, 
or the
distortion in the physical volume occupied by the observed galaxies,
compared to the volume the observer infers, and the volume effect
can be obtained  \cite{YOFIZA09,YOO10,CHLE11,BODU11,JESCHI12}
by following the geodesic path and computing the
physical volume subtended by the observed angle~$d^2\hat n$ and the observed
redshift~$dz$ as
\bear
\de V&=& 3~\dz+\dg+2~{\drr\over\rbar_z}-2\kappa+H{d\over dz}\drr-\al
+\VV_\para~ \nnn
&=&2~\dD+\dz+H{d\over dz}\drr +\varphi+\VV_\para+\pa_\para^2\ga~,
\enar
where the fluctuation in the metric determinant is
$\dg=\al+3\varphi+\Delta\ga$~.
Therefore, the fluctuation in the observed galaxy number density 
$n_g^\up{obs}=\bar n_g(z)(1+\degi)$ is then
\beeq
\degi=b~\de_v-e_z~\dz_v-t_L~\dD+\de V~.
\eneq
The notation convention in \cite{NADUET21} corresponds to
\bear
&&
f_\up{evo}\RA3-e_z~,\Dquad 5s\RA t_L~,\\
&&
V_\up{obs}\RA-(\pa_\para v)_{\obar}~,\Dquad
\Psi\RA\ax~,\Dquad \Phi\RA-\px~.\nn
\enar

\subsection{Scalar Computation}
To facilitate the computation, we choose the conformal Newtonian
gauge ($\beta\equiv\gamma\equiv\chi\equiv0$). Furthermore, we only consider
{\it scalar} perturbations, ignoring the vector and tensor perturbation.
Under the approximation that there exist negligible anisotropic pressure,
two gravitational potentials in the Newtonian gauge are equivalent
$\psi=-\phi$.

The radial distortion~$\drr$ of the source, the distortion~$\dz$ in the 
observed redshift, and the gravitational lensing convergence~$\kappa$
can be computed in the conformal Newtonian gauge as
\bear
\label{ALLEQ}
&&
\drr=-v_{\obar}-{\dz\over\HH_z}+2\int_0^{\rbar_z}d\rbar~\psi
+n_\alpha\delta x^\alpha_{\obar}~,\\
&&
\dz=-\HH v_{\obar}-\left[\pa_\para v+\psi\right]^{z}_{0}
-2\int_0^{\rbar_z} d\rbar~\psi'~,\\
&&
-2\kappa=-2(\pa_\para v)_{\obar}
-{2n_\alpha\delta x^\alpha_{\obar}\over\rbar_z}
-2\int_0^{\rbar_z}d\rbar\left({\rbar_z-\rbar\over \rbar_z \rbar}\right)
\hat\nabla^2~\psi~,\nn
\enar
where all the quantities are in the conformal Newtonian gauge,
the observer position $\bar x^\mu_{\obar}=(\eta_{\obar},0)$
in the background is represented as~$\obar$, and the conformal time
today is
\beeq
\eta_{\obar}=\int_0^\infty{dz\over H(z)}~.
\eneq
The derivative term in the volume effect~$\de V$ can be further 
manipulated as
\bear
H{d\over dz}\drr&=&-{\HH'\over\HH^2}\dz+{1\over\HH}\pa_\para^2 v
-{1\over\HH}\pa_\para\left(-\psi+v'\right)+{\psi'\over \HH}+2\psi \nnn
&=&-{\HH'\over\HH^2}\dz+{1\over\HH}\pa_\para^2 v
+\pa_\para v+{\psi'\over \HH}+2\psi~,\nn
\enar
where in the last equality we used the conservation equation
$v'+\HH v=\psi$.

At the linear order in perturbations, the contributions to the observed
galaxy number density fluctuation~$\degi(z,\Nang)$, the luminosity distance
fluctuation~$\dD(z,\Nang)$, or the observed CMB temperature anisotropies
$\Theta_\ga^\up{obs}(\Nang)$ are composed of
three distinct types: one at the source position, one at the observer position,
and one along the line-of-sight direction, e.g.,
\beeq
\label{types}
\degi=:\de_\up{s}+\de_{\obar}+\de_\up{los}~.
\eneq
The individual terms at the observer position~$\de_{\obar}$
contribute to the angular
monopole and the dipole at the linear order in perturbations (quadrupole
as well, if we consider the tensor perturbations). 
All the terms in~$\de_\up{s}$
and~$\de_\up{los}$ contribute to all angular multipoles, including the
monopole and the dipole.

First, we collect all the terms at the observer position:
\bear
&&
(\dz)_{\obar}=-\HH_0 v_{\obar}+(\pa_\para v)_{\obar}+\psi_{\obar}~,\\
&&
(\drr)_{\obar}=-{1\over\HH_z}\left(\pa_\para v+\psi\right)_{\obar}~,\\
&&
-2(\kappa)_{\obar}=-2(\pa_\para v)_{\obar}~.
\enar
With these expressions, we
obtain
\bear
\label{CMBoo}
&&
(\Theta_\ga^\up{obs})_{\obar}=\HH_0v_{\obar}-\psi_{\obar}-(\pa_\para v)_{\obar}~,
\\
\label{DDoo}
&&
(\dD)_{\obar}
=-\HH_0v_{\obar}\left(1-{1\over\HH_z\rbar_z}+{1\over\HH_0\rbar_z}\right) \\
&&\Tquad
+\psi_{\obar}
\left(1-{1\over\HH_z\rbar_z}\right)-{(\pa_\para v)_{\obar}\over\HH_z\rbar_z}~,
\nnn
&&
\label{obss}
(\degi)_{\obar}=\HH v_{\obar}\left[h_z-3-{2\over\HH_0\rbar_z}
+t_L\left(1-{1\over\HH_z\rbar_z}+{1\over\HH_0\rbar_z}\right)\right] \nnn
&&\Tquad
-(\pa_\para v)_{\obar}\left(h_z-1-{t_L\over\HH_z\rbar_z}\right) \nnn
&&\Tquad
-\psi_{\obar}\left[h_z-3
+t_L\left(1-{1\over\HH_z\rbar_z}\right)\right]~,
\enar
where we defined the time-dependent coefficient \cite{GRSCET20}
\beeq
h(z):=e_z+{2\over\HH_z\rbar_z}+{\HH'_z\over\HH^2_z}~.
\eneq
Note that the full expression for the observed CMB temperature anisotropies
is presented in Eq.~\eqref{CMB}.

Last, the line-of-sight contributions~$\de_\up{los}$ 
and the contributions~$\de_\up{s}$ at the source position 
can be readily read off from the equations above,
and the final expressions are
\begin{widetext}
\bear
&&
\label{CMBSLOS}
(\Theta_\ga^\up{obs})_\up{los}=\int_0^{\rbar_\star}d\rbar~2\psi'~,\Tquad\Tquad
(\Theta_\ga^\up{obs})_\up{s}=\Theta_\star+\psi+(\pa_\para v)~,\\
&&
\label{DDLOS}
(\dD)_\up{los}=-2\left(1-{1\over\HH_z\rbar_z}\right)\int_0^{\rbar_z}d\rbar~\psi'
+{2\over\rbar_z}\int_0^{\rbar_z}d\rbar~\psi
-\int_0^{\rbar_z}d\rbar\left({\rbar_z-\rbar\over \rbar_z \rbar}\right)
\hat\nabla^2~\psi~,\\
&&
\label{DDSS}
(\dD)_\up{s}=
-\pa_\para v\left(1-{1\over\HH_z\rbar_z}\right)-\psi\left(2-{1\over\HH_z\rbar_z}
\right)~,\\
&&
\label{OBSLOS}
(\degi)_\up{los}=\int_0^{\rbar_z}d\rbar\left\{{4-2~t_L\over\rbar_z}~\psi
+\left[2\left(h_z-3\right)+2~t_L\left(1-{1\over\HH_z\rbar_z}\right)\right]\psi'
+(t_L-2)\left({\rbar_z-\rbar\over \rbar_z \rbar}\right)
\hat\nabla^2~\psi\right\}~,\\
&&
\label{OBSSRC}
(\degi)_\up{s}=b~\de_v+{1\over\HH_z}\pa_\para^2v
+\left[h_z-3+t_L\left(1-{1\over\HH_z\rbar_z}\right)\right]\pa_\para v
+\left[h_z-4+t_L\left(2-{1\over\HH_z\rbar_z}\right)\right]\psi
-e_z\HH_z v+{\psi'\over\HH_z}~.~~~~
\enar
\end{widetext}

\subsection{Angular Decomposition and Transfer Functions}
With three distinct types of contributions in Eq.~\eqref{types},
each type can be Fourier decomposed as
\bear
\de_\up{s}&=&\int{d^3k\over(2\pi)^3}~e^{i\kvec\cdot\xvec_z}~\de_\up{s}(k,\rbar_z)
~,\\
\de_{\obar}&=&\int{d^3k\over(2\pi)^3}~\de_{\obar}(k,0)~,\\
\de_\up{los}&=&\int{d^3k\over(2\pi)^3}~\int_0^{\rbar_z}d\rbar~
e^{i\kvec\cdot\xvec_{\rbar}}~\de_\up{los}(k,\rbar)~,
\enar
where the position vector~$\xvec=\rbar \Nang$ and
the contributions~$\de_{\obar}$ at the observer position
are the usual Fourier transformation evaluated
at~$\obar$ (hence the exponential factor is unity).
In each type, there exist numerous perturbation contributions, and we
define the transfer function~$\TT_{\de p}$ to describe these perturbations, 
e.g., 
\beeq
\de p(\kvec,\eta)=:\TT_{\de p}(k,\eta)\RR(\kvec)~,
\eneq
in terms of the comoving-gauge curvature perturbation~$\RR(\kvec)$
at the initial time. 
Note that in expressing the transfer function~$\TT_{\de p}(k,\eta)$
we have interchangeably used the redshift~$z$ or 
the comoving distance~$\rbar$ for the time coordinate~$\eta$
throughout the manuscript.
As the observed galaxy fluctuation~$\de_g$
or the luminosity distance fluctuation~$\dD$  is
angular decomposed in terms of spherical harmonics~$Y_{lm}(\Nang)$ as 
\beeq
\de_g(z,\Nang)=\sum_{lm} a_{lm}(z)Y_{lm}(\Nang)~,
\eneq
the angular multipoles can be obtained as
\begin{widetext}
\beeq
\label{TRALM}
a_{lm}(z)=\int d^2\hat n~Y_{lm}^*(\Nang)\degi(z,\Nang)
=4\pi i^l\int \frac{dk\, k^2}{2\pi^2}  \bigg[\qquad
\underbrace{\TT_{\de p}(k,z)j_l(k\rbar_z)+ \cdots}_{=:\TT_l(k,z)}\qquad
\bigg] \int {d\Omega_k\over 4\pi} ~Y_{lm}^*(\Kang) \RR(\bm{k}) ~,
\eneq
\end{widetext}
where we used the plane-wave expansion 
\beeq
\label{plane}
e^{i\kvec\cdot\xvec}=4\pi \sum_{lm} i^l j_l\left(kr\right) 
Y^*_{lm}(\Kang) Y_{lm}(\Nang)~, 
\eneq
$\TT_{\de p}$ is the transfer function for individual perturbation
contributions, and we defined the full angular multipole transfer
function~$\TT_l(k,z)$ inside the square bracket.
Mind that the transfer function is real.
Given the ensemble average in the initial conditions
\bear
\left\langle\RR(\kvec_1)\RR^*(\kvec_2)\right\rangle&=&
(2\pi)^3\de^D(\kvec_1-\kvec_2)~P_\RR(k_1) \\
&=& {(2\pi)^3\over k^2_1}
P_\RR(k_1)\delta^D(k_1-k_2)\delta^D(\Omega_{k_1}-\Omega_{k_2})~,\nn
\enar
the angular power spectrum can be obtained as
\bear
&&
\left\langle a_{lm}(z_1)a_{l'm'}^*(z_2)\right\rangle
=\de_{ll'}\de_{mm'}C_l(z_1,z_2)~ \\
&&\qquad
= \de_{ll'}\de_{mm'}\times
4\pi\int d \ln k \,  \Delta^2_\RR\left(k\right) \TT_l(k,z_1) \TT_l(k,z_2)
~. \nn
\enar

First, we compute the transfer function~$\TT_l(k,z)$ for the contributions
at the source position. For the various contributions at the source position
in Eqs.~\eqref{CMBSLOS}, \eqref{DDSS}, and~\eqref{OBSSRC}, 
the angular momentum dependence
arises primarily from the exponential in the plane wave expansion in
Eq.~\eqref{plane}. 
However, the contributions with additional dependence on~$\Nang$ have
an intrinsic angular momentum, which needs to be added to the orbital
angular momentum from the plane wave expansion.

The contributions without additional angular dependence take the form
$C_i(z)\de p_i(z,\Nang)$ in Eqs.~\eqref{DDSS} and~\eqref{OBSSRC},
and their contributions to the dipole transfer function can be read off
by using Eq.~\eqref{TRALM} as
\beeq
\TT_1(k,z)\ni C_i(z)\TT_{\de p_i}(k,z)j_1(k\rbar_z)~,
\eneq
where the redshift-dependent coefficients for the observed galaxy
fluctuation are
\bear
&&
C_{\de_v}:=b~,\Dquad C_v:=-e_z\HH_z~,\Dquad C_{\psi'}:={1\over\HH_z}~,\nnn
&&
C_\psi:=h_z-4+t_L\left(2-{1\over\HH_z\rbar_z}\right)~,
\enar
that for the luminosity distance fluctuation is
\beeq
C_\psi^{\dD}:=-2+{1\over\HH_z\rbar_z}~,
\eneq
and those for the observed CMB temperature anisotropies are
\beeq
C_\Theta^\ga:=1~,\Tquad C_\psi^\ga:=1~.
\eneq
We put the superscript~$\dD$ or~$\ga$ to contrast the difference in the
quantities for the luminosity distance fluctuation and the CMB temperature
anisotropies, compared to the quantities for the observed galaxy fluctuation. 

The transfer function for
the line-of-sight peculiar velocity $(\pa_\para v)$ at the source position
can be obtained by acting the line-of-sight derivative for the transfer
function of the $v$-contribution shown in Eq.~\eqref{TRALM}. The derivative
operator~$\pa_\para$ acts on the spherical Bessel function~$j_l(k\rbar_z$) 
to yield
\beeq
\TT_1(k,z)\ni C_{\pa_\para v}(z)~k\TT_v(k,z)j_1'(k\rbar_z)~,
\eneq
where the redshift-dependent coefficients are
\bear
C_{\pa_\para v}&:=&h_z-3+t_L\left(1-{1\over\HH_z\rbar_z}\right)~,\\
C_{\pa_\para v}^{\dD}&:=&-1+{1\over\HH_z\rbar_z}~,\Dquad
C_{\pa_\para v}^{\ga}:=1~.
\enar
With the spherical Bessel function identity
\beeq
j_l'(x)={l\over2l+1}~j_{l-1}(x)-{l+1\over2l+1}~j_{l+1}(x)~,
\eneq
it is clear that the derivative operator brings in additional angular
momentum to the orbital angular momentum of the velocity contribution
at the source position.

In the same way, two derivatives $(\pa^2_\para v)$ of the line-of-sight
peculiar motion at the source position yield the transfer function
\beeq
\TT_1(k,z)\ni {k^2\over\HH_z}\TT_v(k,z)j_1''(k\rbar_z)~,
\eneq
which arises from the redshift-space distortion \cite{KAISE87}.
The luminosity distance fluctuation~$\dD$ or the observed CMB temperature
anisotropies has no such contribution.
Again, the angular momentum addition can be obtained by using
\bear
j_l''(x)&=&{l(l-1)\over(2l+1)(2l-1)}~j_{l-2}(x)
-{2l^2+2l-1\over(2l-1)(2l+3)}~j_l(x) \nnn
&&+{(l+1)(l+2)\over(2l+1)(2l+3)}~j_{l+2}(x)~. \nn
\enar

Second, we consider the line-of-sight contributions to the 
angular multipole transfer function.
These contributions include three possibilities:
the line-of-sight integration of~$\psi$, the line-of-sight integration
of~$\psi'$, or the lensing contribution, which we parametrize as
\beeq
\int_0^{\rbar_z}d\rbar\Big[L_\psi(z)\psi+L_{\psi'}(z)\psi'
+(t_L-2)\left({\rbar_z-\rbar\over \rbar_z \rbar}\right)
\hat\nabla^2~\psi\Big]~.
\eneq
Each component at one point along the line-of-sight direction can be
decomposed exactly in the same way as for the contributions at the source
position, except the lensing contribution, where the angular Laplacian 
operator~$\hat\nabla^2$ acts on the spherical harmonics to yield
\beeq
\hat\nabla^2Y_{lm}(\Nang)=-l(l+1)Y_{lm}(\Nang)~.
\eneq
Altogether,
the angular multipole transfer function from the line-of-sight
contributions in Eqs.~\eqref{CMBSLOS}, \eqref{DDLOS}, and~\eqref{OBSLOS}
is
\begin{widetext}
\beeq
\TT_l(k,z)\ni  \int_0^{\rbar_z}d\rbar
\Big[L_\psi(z)\TT_\psi(k,\rbar)+L_{\psi'}(z)\TT_{\psi'}(k,\rbar)
-l(l+1)(t_L-2)\left({\rbar_z-\rbar\over \rbar_z \rbar}\right)\TT_\psi(k,\rbar)
\Big]j_l(k\rbar)~,
\eneq
\end{widetext}
where the redshift-dependent coefficients are
\bear
L_\psi&=&{4-2t_L\over\rbar_z}~, \\
L_{\psi'}&=&2\left(h_z-3\right)+2t_L\left(1-{1\over\HH_z\rbar_z}\right)~,
\enar
for the observed galaxy fluctuation, and
\beeq
L^{\dD}_\psi={2\over\rbar_z}~,\Dquad
L^{\dD}_{\psi'}=-2\left(1-{1\over\HH_z\rbar_z}\right)~,
\eneq
for the luminosity distance fluctuation,
and $L_{\psi'}^\ga=2$ for the observed CMB temperature anisotropies.

Next, we compute the transfer functions from the contributions at the
observer position. Among all the contributions at the observer position
in Eqs.~\eqref{CMBoo}, \eqref{DDoo}, and~\eqref{obss}, 
only the peculiar motion $(\pa_\para v)_{\obar}$
of the observer possesses the directional dependence~$\Nang$, contributing
to the dipole transfer function, while the remaining terms at the observer
position contribute to the monopole transfer function.
This contribution to the dipole transfer function
can be written in Fourier space as
\beeq
O_{\pa_\para v}(z)(\pa_\para v)_{\obar}=
\int{d^3k\over(2\pi)^3}\Big[O_{\pa_\para v}(z)(i\bm{k}\cdot\bm{n})\TT_{v}(k,
\eta_{\obar})\Big]\RR(\bm{k})~,
\eneq
where the derivative operator acted on the Fourier mode before getting
evaluated at the observer position. The redshift-dependent coefficients
for the observed galaxy fluctuation and the luminosity distance are
\bear
O_{\pa_\para v}&:=&1-h_z+{t_L\over\HH_z\rbar_z}~,\\
O^{\dD}_{\pa_\para v}&:=&-{1\over\HH_z\rbar_z}~,\Dquad O^\ga_{\pa_\para v}=-1~.
\enar
Expressing the unit directional vectors~$\Nang$ and~$\Kang$ in terms
of spherical harmonics,
\begin{widetext}
\beeq
\Nang=(\sin\theta\cos\phi,\sin\theta\sin\phi,\cos\theta)=
\sqrt{4\pi\over3}\left({Y_{1,-1}(\Nang)-Y_{1,1}(\Nang)\over\sqrt2},
~i{Y_{1,-1}(\Nang)+Y_{1,1}(\Nang)  \over\sqrt2},~Y_{10}(\Nang)\right)~,
\eneq
the inner product of two directional vectors is
\beeq
\Kang\cdot\Nang={4\pi\over3}\left[Y_{1,1}(\Nang)Y^*_{1,1}(\Kang)+Y_{1,-1}(\Nang)
Y^*_{1,-1}(\Kang)+Y_{10}(\Nang)Y_{10}(\Kang)\right]~,
\eneq
where we used $Y_{l,-m}=(-1)^mY_{lm}^*$ and $Y_{11}=-Y_{1,-1}^*$ to
simplify the expression. Hence we derive the angular dipole coefficient
\beeq
a_{1m}=4\pi i\int \frac{dk\, k^2}{2\pi^2}  \bigg\{\quad\Big[
O_{\pa_\para v}(z){k\over3}\TT_v(k,\eta_{\obar})\Big]
\Big(\de_{m0}+\de_{m,1}+\de_{m,-1}\Big)\quad\bigg\}
 \int {d\Omega_k\over 4\pi} ~ Y^*_{1m}(\Kang)\RR(\bm{k})~,
\eneq
\end{widetext}
where the dipole transfer function includes the quantities in the curly bracket.
Though the dipole transfer function from the peculiar motion at the observer
motion depends on the value of~$m$, the dipole power spectrum is
averaged over~$m$ as
\beeq
C_1=\left\la|a_{1m}|^2\right\ra={1\over3}\sum_m\left\la|a_{1m}|^2\right\ra~,
\eneq
and each $m$-mode contributes equally to the dipole power, such that
the $m$-dependence can be ignored and the dipole transfer function is
simply the quantity in the square bracket:
\beeq
\TT_1(k,z)\ni O_{\pa_\para v}(z)~{k\over3}\TT_v(k,\eta_{\obar})~.
\eneq
\\

\subsection{Dipole Transfer Functions}
Collecting all the contributions from the source position, the observer 
position, and the line-of-sight direction, we obtain the total angular
dipole transfer function for the observed galaxy fluctuation:
\begin{widetext}
\bear
\TT_1(k,z)&:=&O_{\pa_\para v}(z){k\over3}\TT_v(k,\eta_{\obar})+ 
\int_0^{\rbar_z}d\rbar
\Big[L_\psi(z)\TT_\psi(k,\rbar)+L_{\psi'}(z)\TT_{\psi'}(k,\rbar)
-2(t_L-2)\left({\rbar_z-\rbar\over \rbar_z \rbar}\right)\TT_\psi(k,\rbar)
\Big]j_1(k\rbar)\nnn
&&+\Big[C_\de(z)\TT_\de(k,\rbar_z)+C_v(z)\TT_v(k,\rbar_z)
+C_\psi(z)\TT_\psi(k,\rbar_z)+C_{\psi'}(z)\TT_{\psi'}(k,\rbar_z)
\Big]j_1(k\rbar_z) \nnn
&&
+C_{\pa_\para v}(z)k\TT_v(k,\rbar_z)j_1'(k\rbar_z)
+{k^2\over\HH_z}\TT_v(k,\rbar_z)j_1''(k\rbar_z)~,
\enar
the total dipole transfer function for the luminosity distance
\bear
\TT_1^{\dD}(k,z)&:=&
O^{\dD}_{\pa_\para v}(z){k\over3}\TT_v(k,\eta_{\obar})+ \int_0^{\rbar_z}d\rbar
\Big[L^{\dD}_\psi(z)\TT_\psi(k,\rbar)+L^{\dD}_{\psi'}(z)\TT_{\psi'}(k,\rbar)
+2\left({\rbar_z-\rbar\over \rbar_z \rbar}\right)\TT_\psi(k,\rbar)
\Big]j_1(k\rbar)\nnn
&&+C^{\dD}_\psi(z)\TT_\psi(k,\rbar_z)j_1(k\rbar_z) 
+C^{\dD}_{\pa_\para v}(z)k\TT_v(k,\rbar_z)j_1'(k\rbar_z)~,
\enar
and the total dipole transfer function for the observed CMB temperature
anisotropies
\beeq
\TT_1^\ga(k):=\Big[\TT_{\Theta}(k,\eta_\star)+\TT_{\psi}(k,\eta_\star)\Big]
 j_1(k\rbar_\star)+k\TT_v(k,\eta_\star)j_1'(k\rbar_\star) 
-\frac13k\TT_{v}(k,\eta_{\obar})+\int_0^{\rbar_\star}d\rbar
~2~\TT_{\psi'}(k,\rbar)j_1(k\rbar)~.
\eneq
The redshift-dependent
 coefficients for the observed galaxy fluctuation are 
(put together here again for convenience)
\bear
&&
O_{\pa_\para v}:=1-h_z+{t_L\over\HH_z\rbar_z}~,\Tquad
L_\psi={4-2t_L\over\rbar_z}~, \Tquad
L_{\psi'}=2\left(h_z-3\right)+2t_L\left(1-{1\over\HH_z\rbar_z}\right)~,~~~\\
&&
C_\de:=b~,\Tquad C_v:=-e_z\HH_z~,\Tquad 
C_\psi:=h_z-4+t_L\left(2-{1\over\HH_z\rbar_z}\right)~,\\
&&
C_{\psi'}:={1\over\HH_z}~,\Tquad
C_{\pa_\para v}:=h_z-3+t_L\left(1-{1\over\HH_z\rbar_z}\right)~,\Tquad
h_z=e_z+{2\over\HH_z\rbar_z}+{\HH'_z\over\HH^2_z}~,
\enar
and the redshift-dependent coefficients for the luminosity distance
fluctuation are
\bear
&&
O^{\dD}_{\pa_\para v}:=-{1\over\HH_z\rbar_z}~,\Tquad
L^{\dD}_\psi:={2\over\rbar_z}~, \Tquad
L^{\dD}_{\psi'}:=-2\left(1-{1\over\HH_z\rbar_z}\right)~,\\
&&
C^{\dD}_\psi:=-2+{1\over\HH_z\rbar_z}~,\Tquad
C^{\dD}_{\pa_\para v}:=-1+{1\over\HH_z\rbar_z}~.
\enar
\end{widetext}

\section{Shot Noises}
\label{shotnoise}
In the text we have presented the contributions of cosmological fluctuations
to the dipole power in large-scale structure surveys, but the dipole 
measurements from large-scale structure surveys inevitably involve
measurement noises from various sources such as the detector noise, the
telescope response function, the survey window function, and so on.
Here we consider the shot-noise
contribution to the dipole power due to the discreteness of the sources.

The angular number density of galaxies in Eq.~\eqref{map}
can be angular decomposed as $\ngtwo(\Nang)=\sum A_{lm}Y_{lm}(\Nang)$ 
and its power can be computed as
\begin{widetext}
\beeq
\label{full}
\langle A_{lm}A_{l'm'}^*\rangle=\int d\Omega_1~Y_{lm}^*(\Nang_1)\int d\Omega_2
~Y_{l'm'}(\Nang_2)\int dz_1\left({d\bar V\over dz d\Omega}\right)_1
\int dz_2\left({d\bar V\over dz d\Omega}\right)_2
\bigg[\bar n_1\bar n_2\left(1+\xi_g^\up{obs}\right)
+\bar n_1\de^D(\bm{x}_1-\bm{x}_2)\bigg]~,
\eneq
\end{widetext}
where the subscript~1 and~2 represent the corresponding three-dimensional
positions~$\bm{x}$ in terms of the observed redshift~$z$ and angle~$\Nang$
and~$\de^D(\bm{x})$ is the Dirac delta function. The first term in the
square bracket is the background contribution and the second term
is the perturbation contribution or the two-point correlation function,
computed in the main text. The last term is the shot-noise contribution
arising from the discreteness of individual galaxies in surveys.
In the absence of other non-Gaussian systematic errors, the total
angular power spectrum including noises is diagonal:
\beeq
{\cal C}_l^\up{LSS,tot}:=\AVE{|A_{lm}|^2}=
4\pi\bar\Sigma_g^2~\de_{l0}
+\bar\Sigma_g^2\left(C_l^\up{LSS}
+{1\over\bar\Sigma_g}\right)~,
\eneq
where the expression is valid for~$l\geq0$ and $C_{l=0}^\up{LSS}\neq0$ 
(see \cite{YOMIET19,BAYO21,YOEI25} for discussion of the monopole
fluctuations). 
Since the background contributes only to the monopole power,
we can split the observed angular number density 
$\Sigma_g^\up{obs}(\Vang)=\bar\Sigma_g+\de\Sigma_g(\Vang)$ and 
angular decompose as
\beeq
{\de\Sigma_g(\Vang)\over\bar\Sigma_g}=\sum_{lm}a_{lm}Y_{lm}(\Vang)~.
\eneq
The cosmological dipole power~$C_{l=1}^\up{LSS}=\AVE{|a_{1m}|^2}$
computed  in the main text is defined
without $\bar\Sigma_g^2$ in the angular power spectrum
${\cal C}_{l=1}^\up{LSS,tot}$,  and the shot-noise contribution 
to be compared to $C_1^\up{LSS}$ is $C_l^\up{shot}=1/\bar\Sigma_g$, 
which is the same for all angular multipoles ($l\geq0$).

Galaxy surveys measure the three-dimensional galaxy number 
density $\ngobs(z,\Nang)$, and it can be angular decomposed as
$\ngobs(z,\Nang)=\sum A_{lm}(z)Y_{lm}(\Nang)$ to measure the
angular power spectrum $C_l(z)$ as a function of redshift (see, e.g., 
Figure~\ref{Fig:dipole}). 
The angular power spectrum in this case can be readily
computed by using Eq.~\eqref{full} without the integration over the redshift
as
\beeq
{\cal C}_l^\up{LSS,tot}=4\pi\bar n_1\bar n_2~\de_{l0}+\bar n_1\bar n_2
\left[C_l^\up{LSS}+{1\over\bar n_1}{1\over a^3\rbar_1^2}\de^D(\rbar_1-
\rbar_2)\right]~.
\eneq
Mind that the dimension of the galaxy number density is the inverse of
physical volume and the quantities in the square bracket is dimensionless.
The presence of the Dirac delta function makes the shot-noise contribution
infinite, and the method of using $\ngobs$ for the angular power spectrum
is useful only in theoretical understanding. 
Of course, the redshift range~$\Delta z$ for
computing~$\ngobs$ in practice is never infinitesimally small, and hence
the shot-noise is not infinite either. Nevertheless, the
shot-noise contribution in this case is still very large as 
$\bar\Sigma_g\propto \Delta z$. 

For CMB power spectra,
while the width of the last scattering surface is relatively narrow $\Delta z$,
the number of CMB photons obtained over the mission duration is
sufficiently large enough
to make the shot noise contribution negligible, compared
to the detector noise and the cosmic variance contributions to the
power spectrum.

\bibliography{ms.bbl}

\end{document}